\documentclass{article}

\usepackage[preprint]{neurips_2026}

\usepackage[utf8]{inputenc} 
\usepackage[T1]{fontenc}    
\usepackage{hyperref}       
\usepackage{url}            
\usepackage{booktabs}       
\usepackage{amsfonts}       
\usepackage{nicefrac}       
\usepackage{microtype}      
\usepackage{xcolor}         
\usepackage{algorithm}
\usepackage{algorithmic}
\usepackage{amsmath}
\usepackage{amssymb}
\usepackage{bm}
\usepackage{multirow,multicol}
\usepackage{adjustbox}
\usepackage{makecell}
\usepackage{enumitem}
\usepackage{wrapfig}

\title{Frequency-Domain Regularized Adversarial Alignment for Transferable Attacks against Closed-Source MLLMs}

\author{%
  \textbf{Leitao Yuan}\textsuperscript{1,2}\thanks{Equal contribution. Work done during an internship at Shanghai Artificial Intelligence Laboratory.}\quad
  \textbf{Qinghua Mao}\textsuperscript{1,3}\footnotemark[\value{footnote}]\quad
  \textbf{Daizong Liu}\textsuperscript{4}\thanks{Correspondence to: Daizong Liu <daizongliu@whu.edu.cn>, Dongrui Liu <liudongrui@pjlab.org.cn>.}\quad
  \textbf{Kun Wang}\textsuperscript{5}\quad
  \textbf{Wenjie Wang}\textsuperscript{6}\\
  \textbf{Yan Teng}\textsuperscript{1}\quad
  \textbf{Jing Shao}\textsuperscript{1}\quad
  \textbf{Dongrui Liu}\textsuperscript{1}\footnotemark[\value{footnote}]\\[4pt]
  \textsuperscript{1}Shanghai Artificial Intelligence Laboratory \quad
  \textsuperscript{2}Zhejiang University \\
  \textsuperscript{3}Shanghai Jiao Tong University \quad
  \textsuperscript{4}Wuhan University\\
  \textsuperscript{5}Nanyang Technological University \quad
  \textsuperscript{6}University of Science and Technology of China
}

\begin{document}

\maketitle

\begin{abstract}
    Multimodal large language models (MLLMs) remain vulnerable to transfer-based targeted attacks, where perturbations optimized on open-source surrogate encoders can generalize to closed-source MLLMs. A key challenge for improving adversarial transferability is to effectively capture the intrinsic visual focus shared across different models, such that perturbations align with transferable semantic cues rather than surrogate-specific behaviors. However, existing methods suffer from spatial-domain feature redundancy and surrogate-specific gradient signals, thereby hindering cross-model transferability. In this paper, we propose FRA-Attack, which addresses both challenges from a unified frequency-domain regularization perspective. For feature alignment, a high-pass DCT objective on patch features suppresses redundant global structures and concentrates the loss on the high-frequency band that carries the MLLMs' intrinsic visual focus. For gradient optimization, we introduce Frequency-domain Gradient Regularization (FGR), a \textit{model-agnostic} low-pass regularizer that modulates the surrogate gradient using only the geometric frequency coordinate, \textit{i.e.}, no surrogate-derived statistic is involved, so that FGR is model-agnostic by construction, removing surrogate-specific high-frequency artifacts while preserving transferable low-frequency directions. Together, the two components form a unified frequency-domain treatment of transferability. Extensive experiments on $15$ flagship MLLMs across $7$ vendors show that FRA-Attack achieves superior cross-model transferability, particularly with state-of-the-art performance on GPT-5.4, Claude-Opus-4.6 and Gemini-3-flash. 

\end{abstract}

\section{Introduction}

Multimodal large language models (MLLMs) \cite{liu2023visual, bai2025qwen3, hurst2024gpt, comanici2025gemini} have achieved remarkable performance across a wide range of multimodal tasks, including image captioning, visual question answering (VQA), and visual reasoning. Despite their success, recent studies \cite{liu2024safety,gao2024boosting} have revealed that the vision encoders of MLLMs remain highly vulnerable to adversarial attacks \cite{goodfellow2014explain}, where carefully crafted perturbations can mislead model predictions while remaining imperceptible to humans. In practice, attacks against closed-source MLLMs are typically conducted under a transfer-based black-box setting, where attackers have no access to model parameters. To this end, prior works \cite{ zhao2023evaluating,dong2023robust,guo2024efficient,li2025frustratingly, jia2025adversarial,zhao2026pushing} propose optimizing adversarial examples by aligning visual features between source and target images, including both global representations and fine-grained patch-level semantics, on surrogate models, achieving strong transferability to closed-source MLLMs. 

During the perception of clean and adversarial visual patterns, modern MLLMs generally rely on attention-based visual encoders~\cite{radford2021learning,liu2023visual,bai2025qwen3} that aggregate patch tokens into structured semantic representations, where visual content is selectively attended and integrated across layers~\cite{liu2025frequency,weng2026exploring}. This attention-driven process determines the model's intrinsic visual focus by emphasizing semantically meaningful regions and suppressing irrelevant details. \textit{First}, for adversarial examples to transfer across MLLMs, the perturbations must align with the MLLMs' intrinsic visual focus, such that the induced changes consistently affect how models perceive and interpret visual content. This suggests that transferable attacks must ultimately target the intrinsic visual focus shared across models. \textit{Second}, beyond defining the alignment target, transferability further requires the iterative update itself to follow a model-agnostic direction: even when the loss is aligned with the intrinsic visual focus, surrogate gradients carry directions carved out by the surrogate's own decision boundary that any other encoder would not endorse~\cite{dong2023robust,xiong2022stochastic}. Therefore, a transferable attack must satisfy these two coupled conditions, locking the alignment target onto the intrinsic visual focus while keeping each gradient step on a direction shared across MLLMs.

However, existing approaches fall short on both transferable conditions above: (1) Their general spatial-domain visual feature alignment operates on highly redundant and entangled patch representations~\cite{tang2023defects,tang2025towards,liu2025frequency,weng2026exploring}, which emphasize overlapping global structures and inject noisy optimization signals, making it difficult to isolate the intrinsic visual focus on the target image. (2) Besides, during iterative optimization, surrogate gradients carry model-specific directions that do not correspond to the intrinsic visual focus shared across MLLMs~\cite{xiong2022stochastic,zhang2023transferable,ren2025improving}, causing adversarial perturbations to overfit to the surrogate and degrade cross-model transferability.
These two challenges indicate that improving adversarial transferability requires a unified representation that can simultaneously localize the MLLMs' intrinsic visual focus on the alignment side and supply a model-agnostic prior on the gradient side. From this perspective, we revisit both sides using spectrum tools, where different frequency bands provide a natural decomposition of the underlying signal: (1) On the alignment side, frequency-domain feature representations separate redundant global structures from semantically meaningful variations, exposing the band that concentrates the intrinsic visual focus. (2) On the gradient side, the frequency coordinate of the gradient itself supplies a purely geometric prior that is independent of any surrogate-derived statistic, separating model-agnostic directions from surrogate-specific artifacts. The same frequency decomposition therefore offers a principled foundation for addressing both conditions of cross-model adversarial transferability.

\begin{figure*}[t]
\centering
\newcommand{\panelH}{4.6cm}

\begin{minipage}[t]{0.47\linewidth}
  \centering
  \parbox[c][\panelH][c]{\linewidth}{%
    \centering
    \includegraphics[
    width=\linewidth,
    height=\panelH,
    keepaspectratio,
    ]{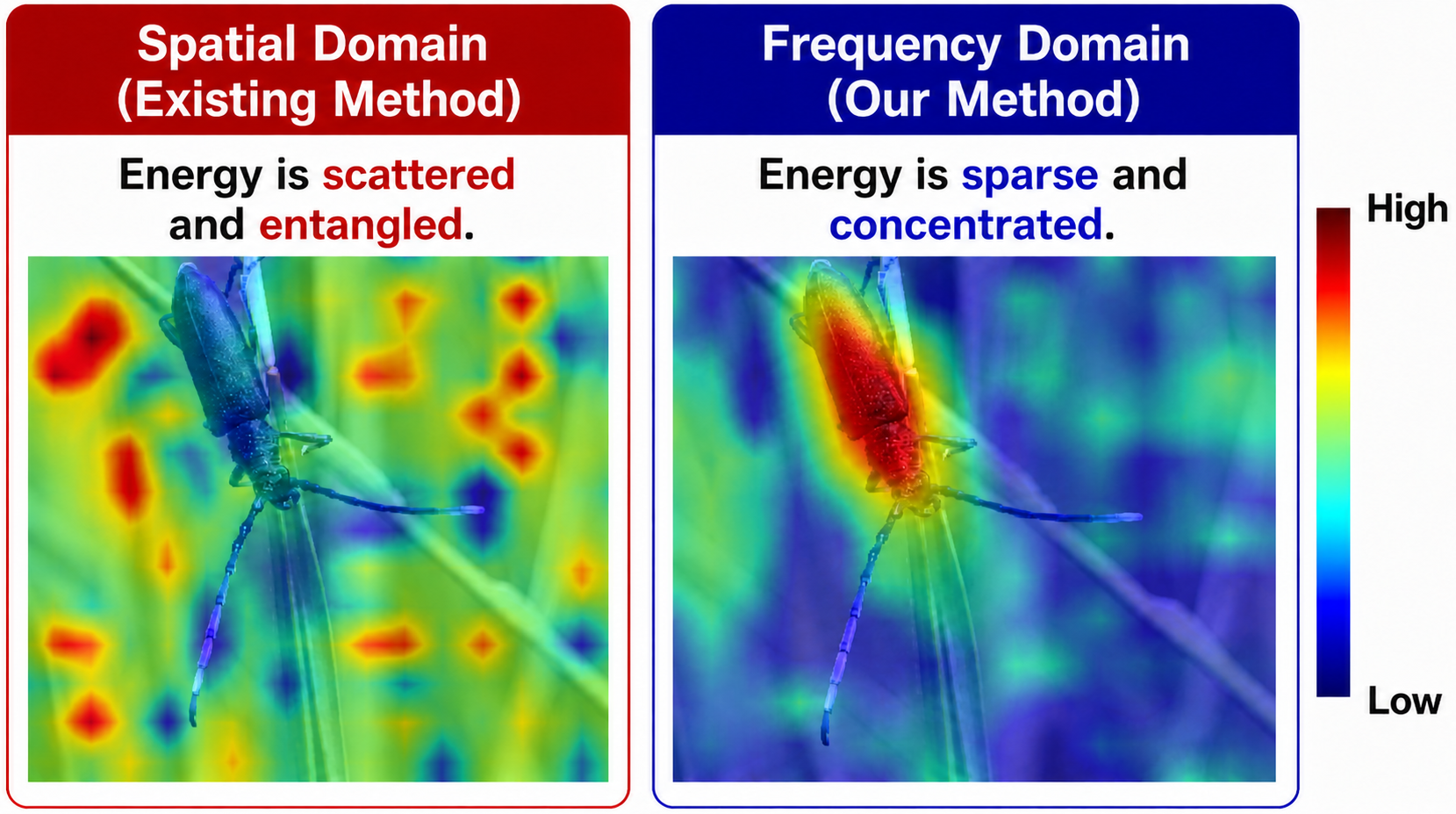}
  }\\[3pt]
  {\small (a) Loss-side: align where it matters.}
\end{minipage}%
\hfill
\begin{minipage}[t]{0.5\linewidth}
  \centering
  \parbox[c][\panelH][c]{\linewidth}{%
    \centering
    \includegraphics[width=\linewidth,height=\panelH,keepaspectratio,]{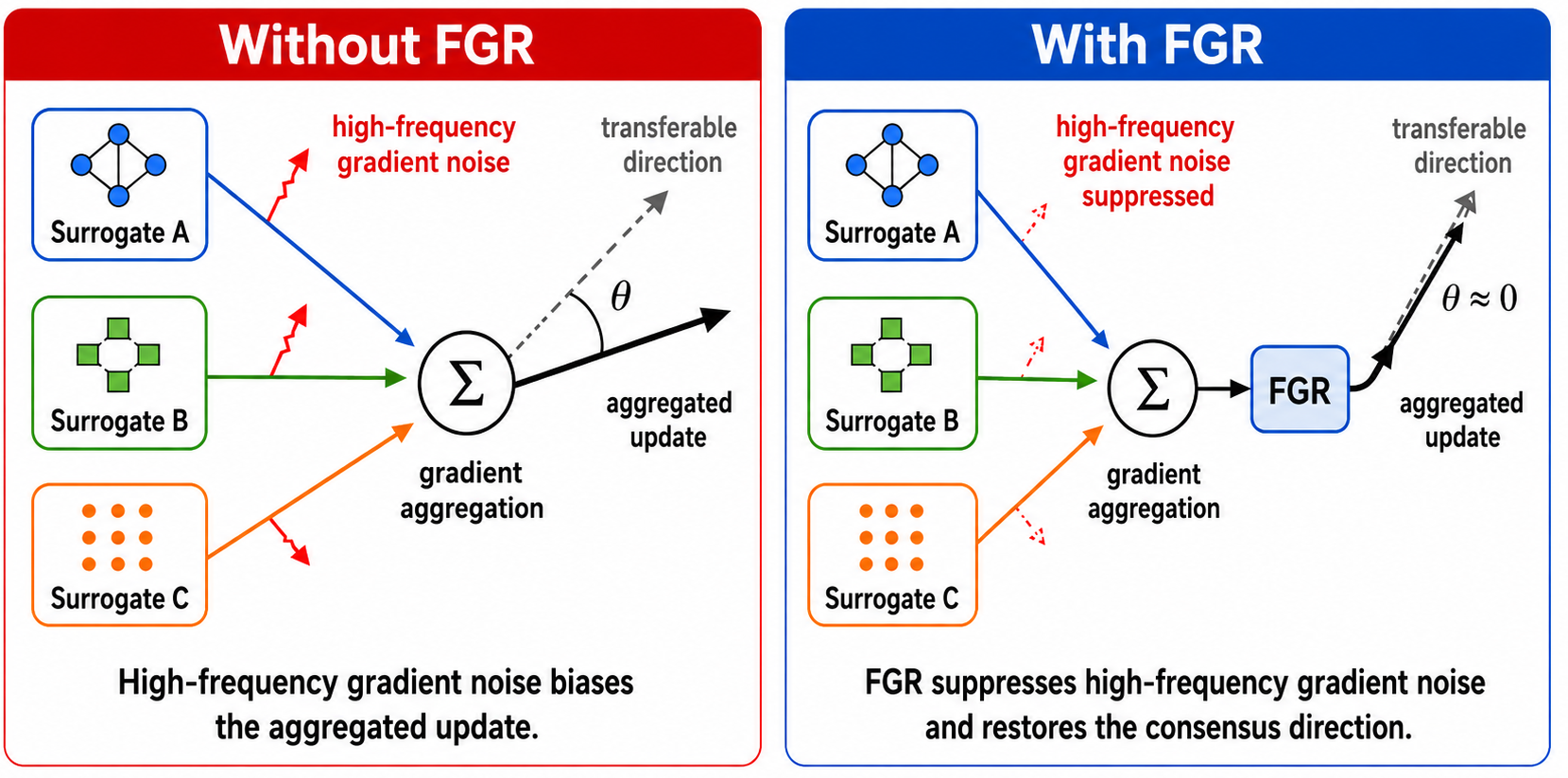}
  }\\[3pt]
  {\small (b) Gradient-side: filter what to follow.}
\end{minipage}

\vspace{-2pt}
\caption{%
\textbf{Two orthogonal frequency-domain levers for transferable VLM attacks.}
(a) \textbf{Loss-side DCT alignment} localizes the transferable visual focus.
(b) \textbf{Gradient-side FGR consensus} suppresses model-specific gradient noise and stabilizes the update direction.
}
\label{fig:teaser}
\vspace{-6pt}
\end{figure*}

Based on the above observations, this work proposes FRA-Attack, a transferable targeted adversarial attack paradigm that improves cross-model transferability by jointly enforcing the two conditions identified above: locking the alignment target onto the intrinsic visual focus shared across MLLMs, and constraining each gradient step to a model-agnostic direction. Rather than directly optimizing surrogate-specific representations or following raw surrogate gradients, FRA-Attack reformulates both sides in a frequency-aware space, where a single frequency decomposition provides a principled tool for addressing both conditions. In particular, FRA-Attack introduces frequency-aware feature alignment on the loss side and frequency-domain gradient regularization on the optimization side, jointly guiding perturbations toward semantically meaningful and transferable directions.

For feature alignment, FRA-Attack moves beyond spatial-domain representations and instead captures visual semantics through a frequency-based decomposition. Spatial patch features are transformed via a Discrete Cosine Transform (DCT)~\cite{ahmed1974discrete}, where high-frequency components emphasize semantically meaningful variations while suppressing redundant global structures, encouraging perturbations to focus on the intrinsic visual focus consistently attended to across models. For gradient optimization, FRA-Attack further introduces a model-agnostic frequency-domain regularization strategy to improve transferability. Instead of directly following surrogate gradients, the gradient is reinterpreted in the frequency domain and selectively modulated according to its transferability: high-frequency components that encode surrogate-specific artifacts are suppressed, while low-frequency components that capture the intrinsic visual focus shared across MLLMs are preserved. This formulation steers adversarial optimization toward directions aligned with the intrinsic visual focus of MLLMs, thereby reducing surrogate overfitting and improving cross-model transferability. 

Our main contributions are summarized as follows:
\begin{itemize}[leftmargin=*,
                itemsep=2pt,
                topsep=2pt,
                parsep=0pt,
                partopsep=0pt]
    \item We propose FRA-Attack, a transfer-based targeted adversarial attack on MLLMs that treats both the characteristics of adversarial transferability of \emph{alignment target} and the \emph{optimization step} in the frequency domain. FRA-Attack addresses two distinct spatial-domain failure modes: redundant patch features on the alignment side, and surrogate-specific noise on the gradient side.
    \item We propose a frequency-domain alignment mechanism that effectively captures the intrinsic visual focus shared across MLLMs, thereby dispelling the spatial-domain redundancy in which the local features merely duplicate the global structures.
    \item We propose Frequency-domain Gradient Regularization (FGR), a model-agnostic gradient regularization strategy that exploits only the geometric frequency coordinate and no surrogate-derived statistic, so that transferable directions are preserved by construction while surrogate-specific noise is removed.
    \item We evaluate FRA-Attack on 15 MLLMs across 7 vendors (OpenAI, Anthropic, Google, Meta, Moonshot, Zhipu, Alibaba) and three open-source vision encoder families (CLIP, AIMv2, SigLIP). FRA-Attack consistently outperforms the strongest prior method across vendors and encoder families, with the largest absolute gain observed on the Anthropic family.
\end{itemize}

\section{Method}

\subsection{Problem Formulation}
\label{sec:prelim}
Given a source image $\bm{x}_s$ and a target image $\bm{x}_t$, we aim to generate an adversarial image $\bm{x}_{adv}$ that is visually indistinguishable from $\bm{x}_s$ yet induces a victim MLLM to produce descriptions semantically aligned with the target image $\bm{x}_t$.
Following~\cite{li2025frustratingly,jia2025adversarial}, we employ an ensemble of $J$ pre-trained vision encoders $\{f_j\}_{j=1}^{J}$ (\textit{e.g.}, CLIP variants) as surrogate models and formulate the attack as a constrained optimization problem:
\vspace{-5pt}
\begin{equation}
  \min_{\bm{x}_{adv}} \sum_{j=1}^{J} W_j \cdot \mathcal{L}_j\!\left(f_j(\bm{x}_{adv}),\; f_j(\bm{x}_t)\right), \quad
  \text{s.t.} \;\; \|\bm{x}_{adv} - \bm{x}_s\|_\infty \le \epsilon,
  \label{eq:formulation}
\end{equation}
where $W_j$ is a per-model weight set adaptively across iterations (Section~\ref{sec:method}) and $\epsilon$ is the perturbation budget.
For each encoder $f_j$, a ViT backbone produces a global feature $\bm{g}_j \in \mathbb{R}^{d_j}$ (the \texttt{[CLS]} token) and a set of patch embeddings $\bm{E}_j \in \mathbb{R}^{P_j \times d_j}$, where $P_j$ is the number of patches and $d_j$ is the embedding dimension of $f_j$.

\noindent\textbf{Victim Models.}
We consider black-box MLLMs (\textit{e.g.}, GPT, Gemini, Claude) as victim models. The adversary has no access to their architectures, parameters, or intermediate representations, and can only query them via APIs to obtain textual responses. Following prior work~\cite{zhao2023evaluating,jia2025adversarial}, we adopt a transfer-based setting, where adversarial perturbations are optimized on open-source surrogate encoders and transferred to the victim models.

\noindent\textbf{Adversary Goals.}
The adversary aims to generate an adversarial example $\bm{x}_{adv}$ that remains visually similar to the source image $\bm{x}_s$, while inducing the victim MLLM to produce outputs semantically aligned with the target image $\bm{x}_t$. The attack is considered successful if the generated response reflects the target semantics rather than the original content.

\subsection{Preliminary Analysis}
\label{sec:baseline}

The dominant recipe for transfer-based attacks~\cite{li2025frustratingly,zhao2026pushing,jia2025adversarial} aligns spatial-domain local features on a surrogate ensemble and optimizes the perturbation along the raw surrogate gradient. Building on the two transferability conditions identified in the introduction, we now formalize where this recipe breaks down on each side: on the alignment side, prior local objectives reduce to the same low-frequency content as the global term and never localize the intrinsic visual focus; on the gradient side, the surrogate-specific component of the input gradient concentrates in a particular frequency band that current remedies do not address.

\noindent\textbf{Limitation 1: Spatial-domain local features do not localize the intrinsic visual focus.}
The intrinsic visual focus lives on a small set of discriminative tokens, yet spatial patch features fail to isolate this set. The multi-scale \texttt{[CLS]} cropping family~\cite{li2025frustratingly,zhao2026pushing} aligns the global token across patch sub-regions, but since neighboring ViT patches are highly correlated~\cite{liu2025frequency,weng2026exploring}, each per-crop \texttt{[CLS]} reduces to a low-frequency aggregate that duplicates rather than complements the global term. The patch-clustering family~\cite{jia2025adversarial} aligns cluster centers $\{\bm{c}_k\}$, which are weighted averages of correlated tokens and inherit the same low-frequency content, with limited capacity for the high-frequency details that carry subject identity. As Fig.~\ref{fig:teaser} shows, spatial-domain energy spreads across background patches while a token-wise frequency decomposition concentrates it on the semantic object. Fine-grained alignment, the very purpose of introducing a local term, is therefore weakened by spatial redundancy, and the intrinsic visual focus is never explicitly targeted.

\noindent\textbf{Limitation 2: Surrogate-specific gradient noise persists without a model-agnostic decomposition.}
Even with a well-chosen alignment target, the surrogate gradient mixes a transferable cross-model component with a surrogate-specific one; prior analysis~\cite{liu2025frequency} localizes the latter predominantly in the high-frequency band of the input gradient spectrum. Existing remedies~\cite{dong2023robust,xiong2022stochastic,zhang2023transferable,ren2025improving} dilute this noise indirectly through input augmentation or model averaging without decomposing the gradient itself, so the surrogate-specific component persists across iterations and is repeatedly accumulated under sign-based updates. What is missing is a principled prior that identifies which gradient components to filter without referring to any surrogate-derived statistic.

The two limitations act on different tensors---patch embeddings and input gradients---and therefore admit two complementary frequency-domain treatments, presented in Section~\ref{sec:overview}.
\vspace{-5pt}

\vspace{-2pt}
\subsection{Frequency-domain Regularized Alignment Attack (FRA-Attack)}
\label{sec:overview}
\label{sec:method}

\begin{figure*}[t]
\centering
\vspace{-18pt}
\includegraphics[width=\linewidth]{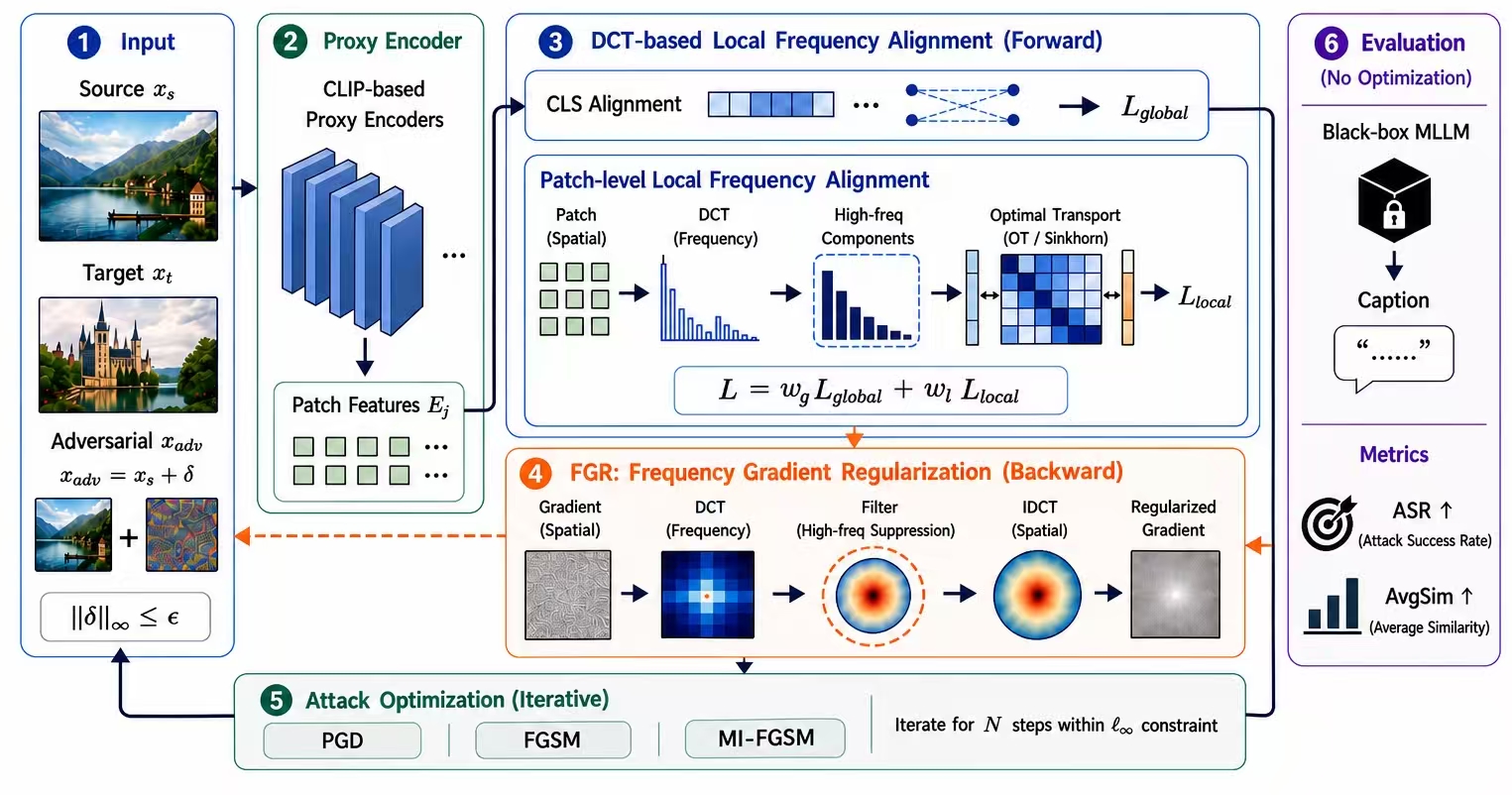}
\vspace{-14pt}
\caption{\textbf{Overall pipeline of FRA-Attack.}
Given a source image $\bm{x}_s$ and a target $\bm{x}_t$, we update the perturbation by (i) aligning DCT high-frequency components of patch embeddings via optimal transport and (ii) regularizing input gradients with a radial low-pass filter.}
\label{fig:pipeline}
\vspace{-12pt}
\end{figure*}

We propose FRA-Attack, a targeted adversarial attack paradigm that improves transferability by treating the two limitations of Section~\ref{sec:baseline} on their respective tensors. The pipeline is illustrated in Fig.~\ref{fig:pipeline} and consists of two components.
The first component (Section~\ref{sec:freq}) replaces the spatial-domain local objective with an optimal transport over high-frequency DCT components of the patch embeddings, locking the alignment onto the intrinsic visual focus rather than the low-frequency content already covered by the global \texttt{[CLS]} term.
The second component (Section~\ref{sec:fgr}) replaces the raw input gradient with a radially low-pass-filtered version before each sign step, suppressing the surrogate-specific noise that accumulates under iterative optimization.

The two components act on different tensors---patch embeddings and input gradients---and therefore combine without interference. We slot them into the standard transfer-attack pipeline of~\cite{jia2025adversarial}: the per-model loss $\mathcal{L}_j$ in Eq.~\eqref{eq:formulation} is the sum of a global \texttt{[CLS]} cosine term and the high-frequency OT term $\mathcal{L}_j^{freq}$ of Section~\ref{sec:freq}; the ensemble weights $W_j$ follow the per-step adaptive scheme of~\cite{jia2025adversarial}; and the FGR-regularized gradient drives an MI-FGSM update. The full equations and the integrated algorithm are deferred to Algorithm~\ref{alg:fra} (Appendix~\ref{app:algorithm}).

\subsection{Frequency-Aware Visual Intrinsic Perception for Explicit Perturbation}
\label{sec:freq}

We construct the local alignment term on the high-frequency DCT components of the patch embeddings, where the intrinsic visual focus identified in Section~\ref{sec:baseline} resides. Concretely, we apply a token-wise DCT to the patch embeddings, isolate the top-energy components in the high-frequency band as a proxy for the intrinsic visual focus, and align them between the source and target images via optimal transport.

\paragraph{Perceiving Visual Focus of Patches using Spectrum Tools.}
\label{sec:dct}

Given the patch embeddings of vision encoder $\bm{E} = [\bm{e}_0, \bm{e}_1, \ldots, \bm{e}_{P-1}]^\top \in \mathbb{R}^{P \times d}$, we apply the Type-II DCT along the token-wise dimension to transform spatial-domain features into frequency-domain coefficients:
\begin{equation}
  \bm{F}_k = \sum_{n=0}^{P-1} \bm{e}_n \cos\!\left[\frac{\pi}{P}\left(n + \tfrac{1}{2}\right)k\right], \quad k = 0, 1, \ldots, P-1,
  \label{eq:dct}
\end{equation}
where $\bm{F}_k \in \mathbb{R}^d$ is the $k$-th frequency component.
Low-frequency components ($k \approx 0$) capture global structure shared across patches, while high-frequency components ($k \gg 0$) encode spatially distinct fine-grained semantics such as textures, edges, and local patterns; we give a formal derivation in Appendix~\ref{app:theory-dct}.

The DCT representation has two properties that make it suitable for iterative adversarial optimization. First, it concentrates the signal energy into a small number of frequency components, yielding a compact representation, as visualized in the right panel of Fig.~\ref{fig:teaser}. Second, the decomposition is \emph{deterministic} and \emph{orthogonal}, yielding a stable and reproducible representation: identical inputs induce identical frequency coefficients, in contrast to K-means cluster centers that are sensitive to initialization and may drift across iterations.

\paragraph{Frequency-Domain Feature Alignment.}
\label{sec:highfreq}

Following the local-alignment recipe of prior transfer-based attacks~\cite{li2025frustratingly,jia2025adversarial}, we align a small set of source and target features, but draw them from the high-frequency DCT band rather than from spatial cluster centers. Concretely, we define a frequency threshold $\theta$ that separates the low- and high-frequency bands of the DCT spectrum, and select the top-$n$ most energetic components from the high-frequency band:
\begin{equation}
  \varepsilon_k = \|\bm{F}_k\|_2, \quad k = 0, \ldots, P{-}1,
  \label{eq:energy}
\end{equation}
\begin{equation}
  \mathcal{I} = \operatorname{Top\text{-}n}\!\left(\{\varepsilon_k\}_{k=\theta}^{P-1}\right),
  \label{eq:topn}
\end{equation}
where $\operatorname{Top\text{-}n}(\cdot)$ returns the indices of the $n$ largest values. The selected high-frequency features are:
\begin{equation}
  \bm{L} = [\bm{F}_{i_1},\; \bm{F}_{i_2},\; \ldots,\; \bm{F}_{i_n}]^\top \in \mathbb{R}^{n \times d}, \quad \{i_1, \ldots, i_n\} = \mathcal{I}.
  \label{eq:local_feat}
\end{equation}

Given the high-frequency features $\bm{L}^{(s)}, \bm{L}^{(t)} \in \mathbb{R}^{n \times d}$ from the adversarial and target images, we align them using an entropic optimal transport formulation. The cost matrix $\bm{C}_{ab}$ is defined as the cosine distance between $\ell_2$-normalized features, with uniform marginals $\tfrac{1}{n}\bm{1}_n$. The resulting alignment loss is:
\begin{equation}
  \mathcal{L}_{freq} = \sum_{a,b} \bm{C}_{ab}\,\pi^*_{ab},
  \label{eq:lfreq}
\end{equation}
where $\bm{\pi}^*$ is the Sinkhorn transport plan with entropic regularization parameter $\lambda$.

\subsection{Model-Agnostic Gradient Regularization for Adversarial Transferability}
\label{sec:fgr}


We construct the low-frequency selector on input gradients, realizing the second component outlined in Section~\ref{sec:overview}. The goal is to remove surrogate-specific noise from the input gradient before each sign update, rather than augmenting the input pixels as in prior spectrum-based attacks~\cite{dong2023robust}.

We propose \emph{Frequency-domain Gradient Regularization} (FGR), a model-agnostic mechanism that operates directly on the input gradient. FGR maps the gradient $\bm{\nabla} \in \mathbb{R}^{B \times C \times H \times W}$ to the spatial frequency domain via a 2D Type-II DCT, applies a frequency-dependent attenuation based on the radial distance from the DC component, and transforms the result back with the inverse DCT. The radial distance is defined as
\begin{equation}
  d(u, v) = \frac{\sqrt{u^2 + v^2}}{\sqrt{H^2 + W^2}} \in [0, 1],
  \label{eq:freq_dist}
\end{equation}
where $d=0$ corresponds to the DC component and $d=1$ to the highest frequency. The regularized gradient is
\begin{equation}
  \tilde{\bm{\nabla}} = \operatorname{IDCT}_{\mathrm{2D}}\!\left(\phi(d)\cdot\operatorname{DCT}_{\mathrm{2D}}(\bm{\nabla})\right), \qquad
  \phi(d) = (1 - d)^{p},
  \label{eq:fgr_decay}
\end{equation}
where $\phi$ preserves low-frequency components and attenuates higher-frequency ones. The regularized gradient $\tilde{\bm{\nabla}}$ replaces $\bm{\nabla}$ in any first-order adversarial update (e.g., FGSM, MI-FGSM, PGD), without requiring access to surrogate-internal representations.

\section{Experiments}

\subsection{Experimental Setup}\label{sec:settings}

\noindent\textbf{Setup.}
We use $1{,}000$ source--target image pairs (NIPS~2017 sources resized to $224\times224$, MSCOCO~\cite{lin2014microsoft} validation targets), reused across all methods and victims. The surrogate is a three-CLIP~\cite{radford2021learning} ensemble of ViT-B/16, ViT-B/32, and ViT-g-14-laion2B. The victim panel covers $15$ MLLMs in three groups --- closed-source standard, closed-source thinking (reasoning-augmented variants), and open-source models from six vendors. We compare against four weak baselines~\cite{zhao2023evaluating,guo2024efficient,dong2023robust,zhang2025anyattack} on the first $100$ pairs (marked $^{\dagger}$) and three strong baselines: M-Attack~\cite{li2025frustratingly}, FOA-Attack~\cite{jia2025adversarial}, and M-Attack-V2~\cite{zhao2026pushing}. To keep consistent with the standard image-to-image attack setting, we run M-Attack-V2 with target-side retrieval disabled (it otherwise pulls in an external MSCOCO pool to enlarge each target into an image-to-images set); full details in Appendix~\ref{app:setup-details}. All attacks share perturbation budget $\epsilon = 16/255$ ($\ell_\infty$) and $N = 300$ MI-FGSM iterations. Full victim names, hyperparameters, API endpoints, and hardware are reported in Appendix~\ref{app:setup-details}.

\noindent\textbf{Metrics.}
Following the LLM-as-a-judge protocol of M-Attack~\cite{li2025frustratingly}, the same victim captions both the adversarial and the target images, and a GPT-4o judge measures their similarity. We report \emph{ASR} (fraction with similarity $>0.5$), \emph{AvgSim} (mean similarity), and a complementary \emph{KMR} at three strictness levels $\alpha\!<\!\beta\!<\!\gamma$.

\vspace{-5pt}
\subsection{Main Results}\label{sec:main_results}
\vspace{-5pt}
Across $1{,}000$ source--target pairs, FRA-Attack consistently achieves the strongest performance on all evaluated closed-source MLLMs (Tables~\ref{tab:closed}--\ref{tab:thinking}). Compared with the strongest prior baseline, M-Attack-V2~\cite{zhao2026pushing}, the improvement is particularly notable on the Anthropic family, with ASR improvement of $12.3\%$ on Claude-Opus-4.6 and $10.0\%$ on Claude-Sonnet-4.6. FRA-Attack also generalizes well across heterogeneous open-source vision encoders, ranking first on $5$ of $6$ open-source victims spanning CLIP-style, AIMv2, and SigLIP/SigLIP-2 encoder families (Table~\ref{tab:open}).
\vspace{-5pt}
\paragraph{Strong transferability on commercial closed-source MLLMs.}
FRA-Attack consistently improves over previous transfer-based attacks across all closed-source victims. The improvement is most pronounced on the Anthropic models, while more moderate but still consistent gains are observed on OpenAI and Gemini models. Although the internal visual architectures of these systems are not publicly available, the results suggest that the proposed frequency-domain alignment better captures visual cues that remain stable across different MLLMs. In particular, DCT-based feature alignment encourages perturbations to focus on semantically meaningful regions, while FGR suppresses surrogate-specific optimization noise, together leading to more transferable adversarial directions.

\begin{table}[t]
\centering
\caption{Performance of ASR (\%) and AvgSim on different closed-source MLLMs. }
\label{tab:closed}
\begin{adjustbox}{width=\linewidth}
\scriptsize
\begin{tabular}{l|cc|cc|cc|cc|cc|cc}
\toprule
\multirow{2}{*}{\textbf{Method}}
& \multicolumn{2}{c|}{GPT-5.2}
& \multicolumn{2}{c|}{GPT-5.4}
& \multicolumn{2}{c|}{Claude-Opus-4.6}
& \multicolumn{2}{c|}{Claude-Sonnet-4.6}
& \multicolumn{2}{c|}{Gemini-3-flash}
& \multicolumn{2}{c}{Gemini-2.5-flash} \\
\cmidrule{2-13}
& ASR & AvgSim & ASR & AvgSim & ASR & AvgSim & ASR & AvgSim & ASR & AvgSim & ASR & AvgSim \\
\midrule
AttackVLM$^{\dagger}$~\cite{zhao2023evaluating}    & 1.0 & 0.060 & 0.0  & 0.044 & 0.0  & 0.032 & 0.0 & 0.038 & 0.0  & 0.025 & 0.0 & 0.033 \\
AdvDiffVLM$^{\dagger}$~\cite{guo2024efficient} & 1.0 & 0.047 & 0.0  & 0.036 & 0.0  & 0.035 & 1.0 & 0.043 & 0.0  & 0.054 & 1.0 & 0.046  \\
SSA-CWA$^{\dagger}$~\cite{dong2023robust}          & 0.0 & 0.049 & 0.0  & 0.033 & 0.0  & 0.022 & 0.0 & 0.022 & 0.0  & 0.028 & 0.0 & 0.029 \\
AnyAttack$^{\dagger}$~\cite{zhang2025anyattack}   & 1.0 & 0.053 & 0.0  & 0.057 & 0.0  & 0.039 & 1.0 & 0.039 & 0.0  & 0.038 & 0.0 & 0.043 \\
\midrule
M-Attack~\cite{li2025frustratingly}             & 24.0 & 0.272 & 23.4 & 0.268 & 49.2 & 0.433 & 42.1 & 0.402 & 26.8 & 0.310 & 42.9 & 0.407 \\
FOA-Attack~\cite{jia2025adversarial}              & 30.9 & 0.312 & 27.3 & 0.300 & 51.3 & 0.455 & 49.2 & 0.444 & 30.9 & 0.341 & 47.5 & 0.439 \\
M-Attack-V2 ~\cite{zhao2026pushing}           & 41.2 & 0.383 & 37.9 & 0.369 & 64.5 & 0.536 & 62.9 & 0.532 & 47.5 & 0.437 & 65.9 & 0.544  \\
\midrule
\textbf{FRA-Attack (Ours)}                          & \textbf{46.9} & \textbf{0.427} & \textbf{44.5} & \textbf{0.411} & \textbf{76.8} & \textbf{0.613} & \textbf{72.9} & \textbf{0.595} & \textbf{50.8} & \textbf{0.456} & \textbf{67.4} & \textbf{0.545}   \\
\bottomrule
\end{tabular}
\end{adjustbox}
\vspace{-10pt}
\end{table}

\begin{table}[t]
\centering
\caption{Performance of ASR (\%) and AvgSim on  different closed-source thinking MLLMs. }
\label{tab:thinking}
\begin{adjustbox}{width=0.78\linewidth}
\scriptsize
\begin{tabular}{l|cc|cc|cc}
\toprule
\multirow{2}{*}{\textbf{Method}}
& \multicolumn{2}{c|}{GPT-5.4-thinking}
& \multicolumn{2}{c|}{Claude-Opus-4.6-thinking}
& \multicolumn{2}{c}{Gemini-3-flash-thinking} \\
\cmidrule(lr){2-3}\cmidrule(lr){4-5}\cmidrule(lr){6-7}
 & ASR & AvgSim & ASR & AvgSim & ASR & AvgSim \\
\midrule
M-Attack~\cite{li2025frustratingly}          & 23.0 & 0.264 & 45.6 & 0.414 & 26.8 & 0.306 \\
FOA-Attack~\cite{jia2025adversarial}         & 27.7 & 0.302 & 51.1 & 0.453 & 34.7 & 0.348 \\
M-Attack-V2  ~\cite{zhao2026pushing}    & 38.8 & 0.377 & 66.7 & 0.547 & 45.4 & 0.426 \\
\midrule
\textbf{FRA-Attack (Ours)}                   & \textbf{46.1} & \textbf{0.416} & \textbf{74.0} & \textbf{0.614} & \textbf{51.0} & \textbf{0.456} \\
\bottomrule
\end{tabular}
\end{adjustbox}
\vspace{-8pt}
\end{table}

\vspace{-5pt}
\paragraph{Reasoning-enhanced MLLMs remain vulnerable.}
The advantage of FRA-Attack persists on all thinking-enabled variants in Table~\ref{tab:thinking}, where the performance gap over M-Attack-V2 remains consistently large. Interestingly, enabling chain-of-thought reasoning only marginally changes the ASR for all methods. This observation suggests that the attack primarily affects the visual perception stage before reasoning is performed, such that additional reasoning computation cannot recover from corrupted visual evidence.
\vspace{-5pt}
\paragraph{Transferability across heterogeneous visual encoders.}
FRA-Attack also transfers effectively across diverse open-source encoder families, including CLIP-style, AIMv2, and SigLIP/SigLIP-2 based MLLMs. The improvement over M-Attack-V2 remains consistent across most models, indicating that the proposed frequency-domain formulation is not tied to a specific encoder architecture. We attribute this robustness to the complementary roles of the two components: frequency-domain alignment encourages perturbations to target intrinsic visual focus shared across attention-based encoders, while FGR reduces surrogate-specific high-frequency artifacts that otherwise limit cross-architecture transferability. The only exception is Gemma-3-27B-it, where all methods exhibit relatively low ASR. We discuss this behavior further in Appendix~\ref{app:limits}.

\begin{table}[t]
\centering
\caption{Performance of ASR (\%) and AvgSim on different open-source MLLMs.}
\label{tab:open}
\begin{adjustbox}{width=\linewidth}
\scriptsize
\begin{tabular}{l|cc|cc|cc|cc|cc|cc}
\toprule
\multirow{2}{*}{\textbf{Method}}
& \multicolumn{2}{c|}{Qwen3-VL-8B}
& \multicolumn{2}{c|}{Qwen3-VL-32B}
& \multicolumn{2}{c|}{GLM-4.6V}
& \multicolumn{2}{c|}{Llama-3.2-11B-V}
& \multicolumn{2}{c|}{Gemma-3-27B-it}
& \multicolumn{2}{c}{Kimi-K2.5} \\
\cmidrule(lr){2-3}\cmidrule(lr){4-5}\cmidrule(lr){6-7}\cmidrule(lr){8-9}\cmidrule(lr){10-11}\cmidrule(lr){12-13}
& ASR & AvgSim & ASR & AvgSim & ASR & AvgSim & ASR & AvgSim & ASR & AvgSim & ASR & AvgSim \\
\midrule
AttackVLM$^{\dagger}$~\cite{zhao2023evaluating}   & 1.0 & 0.073 & 1.0 & 0.058 & 0.0 & 0.035 & 1.0 & 0.025 & 16.0 & 0.258 & 0.0 & 0.027 \\
AdvDiffVLM$^{\dagger}$~\cite{guo2024efficient}    & 1.0 & 0.033 & 0.0 & 0.046 & 1.0 & 0.045 & 0.0 & 0.034 & 11.0 & 0.266 & 1.0 & 0.038 \\
SSA-CWA$^{\dagger}$~\cite{dong2023robust}         & 0.0 & 0.024 & 1.0 & 0.043 & 1.0 & 0.033 & 0.0 & 0.025 & 14.0 & 0.272 & 0.0 & 0.017 \\
AnyAttack$^{\dagger}$~\cite{zhang2025anyattack}   & 1.0 & 0.037 & 0.0 & 0.052 & 0.0 & 0.040 & 0.0 & 0.028 & 9.0  & 0.216 & 2.0 & 0.026 \\
\midrule
M-Attack~\cite{li2025frustratingly}                & 59.6 & 0.509 & 52.6 & 0.459 & 62.2 & 0.518 & 36.8 & 0.354 & 18.9 & 0.318 & 39.4 & 0.364 \\
FOA-Attack~\cite{jia2025adversarial}               & 66.3 & 0.547 & 58.2 & 0.491 & 67.7 & 0.554 & 41.7 & 0.380 & 22.1 & 0.339 & 43.6 & 0.408 \\
M-Attack-V2  ~\cite{zhao2026pushing}          & 77.0 & 0.620 & 69.6 & 0.567 & 77.9 & 0.618 & 52.3 & 0.449 & \textbf{30.2} & \textbf{0.396} & 56.3 & 0.483 \\
\midrule
\textbf{FRA-Attack (Ours)}                         & \textbf{80.6} & \textbf{0.653} & \textbf{77.6} & \textbf{0.617} & \textbf{87.1} & \textbf{0.687} & \textbf{57.3} & \textbf{0.482} & 28.0 & 0.369 & \textbf{69.0} & \textbf{0.568} \\
\bottomrule
\end{tabular}
\end{adjustbox}
\vspace{-10pt}
\end{table}

\vspace{-5pt}
\subsection{Ablation Study}\label{sec:ablation_study}

\begin{figure}[t]
\centering
\includegraphics[width=0.78\linewidth]{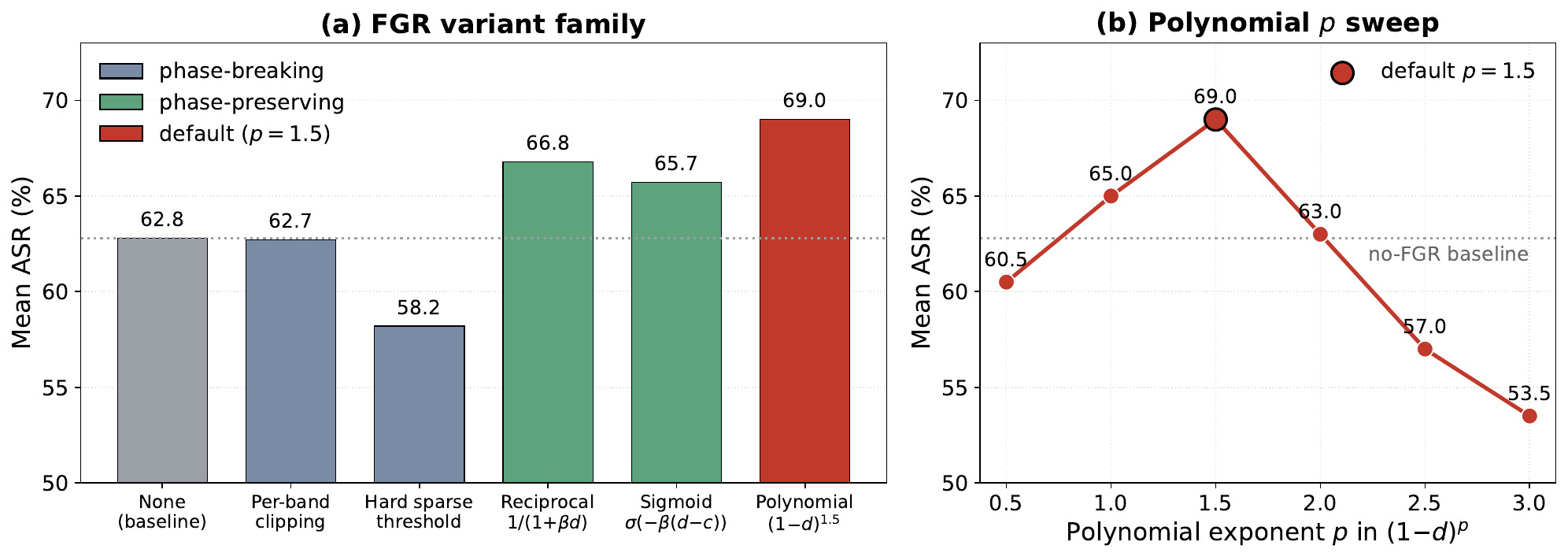}
\vspace{-10pt}
\caption{Ablation Study of FGR design. (a) Mean ASR (\%) for five FGR variants and a no-FGR baseline; (b) Mean ASR (\%) with different Polynomial exponent $p$}
\label{fig:fgr-design}
\vspace{-15pt}
\end{figure}

\begin{wrapfigure}{t}{0.45\linewidth}
\vspace{-15pt}
\centering
\includegraphics[width=\linewidth]{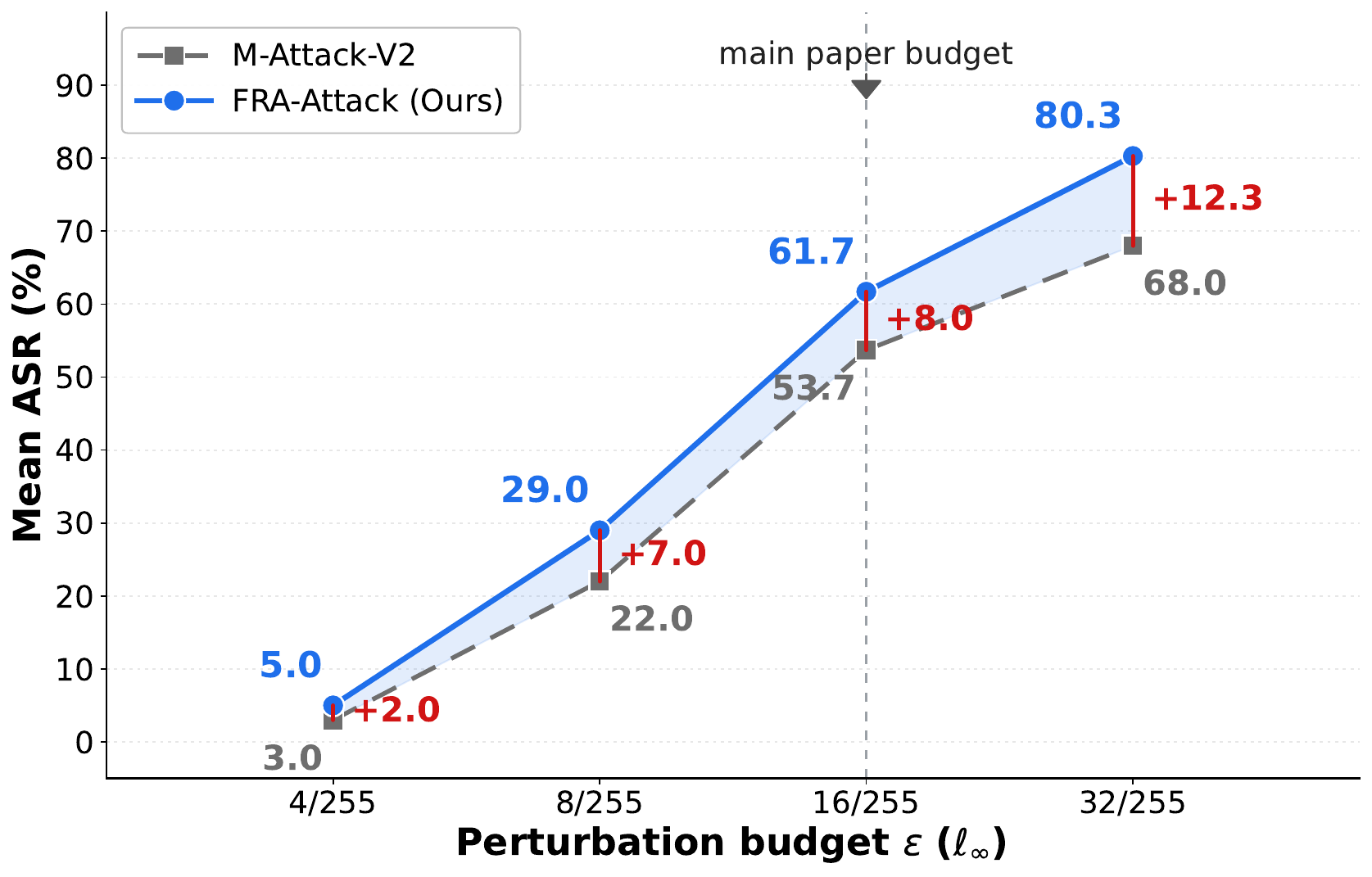}
\vspace{-15pt}
\caption{Mean ASR (\%) of closed-source models with varied $\ell_\infty$ budget.}
\label{fig:eps-sweep}
\vspace{-10pt}
\end{wrapfigure}

Table~\ref{tab:ablation} decomposes FRA-Attack into its two main components. Overall, both DCT alignment and FGR contribute consistently to the final performance, and combining them yields the strongest transferability across all evaluated victims. We further analyze several design choices, including the FGR decay function (Figure~\ref{fig:fgr-design}), optimizer selection (Table~\ref{tab:optimizer}), perturbation budget sensitivity (Figure~\ref{fig:eps-sweep}), and the effect of removing the surrogate ensemble (Table~\ref{tab:single-surrogate}). Additional hyperparameter studies are provided in Appendix~\ref{app:ablation-extended}.

\begin{table}[t]
\centering

\begin{minipage}[t]{0.46\linewidth}
\centering
\caption{Component ablation of FRA-Attack's performance on three closed-source models.}
\label{tab:ablation}
\begin{adjustbox}{width=\linewidth}
\small
\begin{tabular}{l|cc|cc|cc}
\toprule
\multirow{2}{*}{\textbf{Variant}} 
& \multicolumn{2}{c|}{GPT-5.4}
& \multicolumn{2}{c|}{Claude-Opus-4.6}
& \multicolumn{2}{c}{Gemini-3-flash} \\
\cmidrule(lr){2-3}\cmidrule(lr){4-5}\cmidrule(lr){6-7}
& ASR & AvgSim
& ASR & AvgSim
& ASR & AvgSim \\
\midrule
M-Attack~\cite{li2025frustratingly}                            
& 23.4 & 0.268 
& 49.2 & 0.433 
& 26.8 & 0.310 \\

FOA-Attack~\cite{jia2025adversarial}                           
& 27.3 & 0.300 
& 51.3 & 0.455 
& 30.9 & 0.341 \\

M-Attack-V2~\cite{zhao2026pushing}                      
& 37.9 & 0.369 
& 64.5 & 0.536 
& 47.5 & 0.437 \\
\midrule

\textbf{FRA-Attack}                                     
& \textbf{44.5} & \textbf{0.411} 
& \textbf{76.8} & \textbf{0.613} 
& 50.8 & 0.456 \\

\quad w/o FGR   
& 39.0 & 0.387 
& 72.7 & 0.585 
& \textbf{53.1} & \textbf{0.467} \\

\quad w/o DCT alignment
& 38.7 & 0.366 
& 74.0 & 0.597 
& 48.1 & 0.430 \\
\bottomrule
\end{tabular}
\end{adjustbox}
\end{minipage}
\hfill
\begin{minipage}[t]{0.46\linewidth}
\centering
\caption{Replacing the three-CLIP ensemble with a single ViT-B/16 surrogate.}
\label{tab:single-surrogate}
\begin{adjustbox}{width=\linewidth}
\scriptsize
\setlength{\tabcolsep}{2pt}
\begin{tabular}{l|cc|cc|cc}
\toprule
\multirow{2}{*}{\textbf{Method}} 
& \multicolumn{2}{c|}{GPT-5.4}
& \multicolumn{2}{c|}{Claude}
& \multicolumn{2}{c}{Gemini} \\
\cmidrule(lr){2-3}\cmidrule(lr){4-5}\cmidrule(lr){6-7}
& ASR & AvgSim
& ASR & AvgSim
& ASR & AvgSim \\
\midrule
M-Attack~\cite{li2025frustratingly}                     
& 2.0 & 0.080 
& 1.0 & 0.073 
& 2.0 & 0.081 \\

FOA-Attack~\cite{jia2025adversarial}                    
& 3.0 & 0.086 
& 7.0 & 0.139 
& 5.0 & 0.119 \\

M-Attack-V2~\cite{zhao2026pushing}                      
& \textbf{7.0} & 0.138 
& 13.0 & 0.164 
& 7.0 & 0.153 \\

\textbf{FRA-Attack}                              
& 6.0 & \textbf{0.143} 
& \textbf{23.0} & \textbf{0.257} 
& \textbf{14.0} & \textbf{0.205} \\
\bottomrule
\end{tabular}
\end{adjustbox}
\end{minipage}

\vspace{-20pt}
\end{table}







\noindent\textbf{DCT alignment and FGR provide complementary gains.}
Both components consistently improve transferability over M-Attack-V2, and combining them yields the strongest overall performance. DCT alignment improves semantic consistency by focusing perturbations on compact frequency-domain cues, while FGR stabilizes optimization by suppressing surrogate-specific gradient artifacts. The largest gains are again observed on the Anthropic models, where either component alone already recovers most of the final improvement. In contrast, Gemini-3-flash benefits primarily from DCT alignment, while adding FGR provides only marginal additional gain, suggesting a weaker mismatch between surrogate gradients and Gemini's visual pipeline.




\begin{wraptable}{t}{0.5\linewidth}
\vspace{-15pt}
\centering
\caption{ASR (\%) across closed-source models with different optimizer.}
\label{tab:optimizer}

\begin{adjustbox}{width=0.85\linewidth}
\small
\begin{tabular}{l|ccc|c}
\toprule
\textbf{Optimizer} & GPT-5.4 & Claude-Opus-4.6 & Gemini-3-flash & \textbf{Mean ASR} \\
\midrule
FGSM (no momentum)
& 33.0 & 56.0 & 25.0 & 38.0 \\

\textbf{MI-FGSM} ($\mu{=}1.0$)
& 43.0 & \textbf{82.0} & \textbf{55.0} & \textbf{60.0} \\

PGD-Adam
& \textbf{45.0} & 74.0 & 51.0 & 56.7 \\
\bottomrule
\end{tabular}
\end{adjustbox}

\vspace{-4pt}
\end{wraptable}

\noindent\textbf{Smooth frequency-domain regularization is important for transferability.}
Figure~\ref{fig:fgr-design} compares several variants of FGR and shows that continuous radial decay consistently outperforms hard thresholding or statistical clipping. In practice, preserving the relative structure of frequency components is important for maintaining transferable optimization directions. Among all polynomial decay settings, $p{=}1.5$ achieves the best balance between suppressing surrogate-specific noise and retaining transferable semantic signals.

\noindent\textbf{Momentum-based optimization remains critical for transfer attacks.}
Table~\ref{tab:optimizer} shows that optimizer choice has a substantial impact on transferability. Removing momentum leads to a significant drop in ASR, confirming that iterative gradient accumulation is important for stabilizing transferable perturbation directions. Although PGD-Adam performs competitively and slightly improves GPT-5.4 performance, its adaptive normalization reduces the overall cross-model average. We therefore adopt MI-FGSM as the default optimizer throughout the paper.

\noindent\textbf{The advantage of FRA-Attack persists across perturbation budgets.}
Figure~\ref{fig:eps-sweep} evaluates transferability under different $\ell_\infty$ budgets. FRA-Attack consistently outperforms previous methods across all budgets, and the gap becomes more pronounced as the perturbation budget increases. This suggests that gains does not rely on a specific attack strength, but comes from learning more transferable perturbation directions.

\noindent\textbf{FGR remains effective even without surrogate ensembling.}
Table~\ref{tab:single-surrogate} replaces the three-model surrogate ensemble with a single ViT-B/16 encoder. Although all methods experience a substantial performance drop, FRA-Attack still achieves the strongest overall transferability on most victims. The improvement is particularly clear on Claude-Opus-4.6 and Gemini-3-flash, indicating that frequency-domain alignment and gradient regularization remain effective even when cross-model diversity from surrogate ensembling is limited. Results suggest that FGR acts as a transferability-oriented regularizer rather than relying on ensemble-specific effects.





\vspace{-10pt}
\section{Related Works}
\vspace{-8pt}
\subsection{Multimodal Large Language Models}
\vspace{-5pt}
Large language models (LLMs) have demonstrated remarkable performance in numerous applications. Exploiting the potential of LLMs, transformer-based multimodal large language models (MLLMs) can enhance the vision-language semantics from large-scale image-text pairs, facilitating downstream tasks including image captioning, visual question answering, etc. Recent advancements in MLLMs have yielded a diverse ecosystem. Open-source models, including BLIP-2~\cite{li2023blip}, Flamingo~\cite{alayrac2022flamingo}, LLaVA~\cite{liu2023visual}, Llama~3~\cite{grattafiori2024llama}, Gemma~3~\cite{team2025gemma3}, Qwen3-VL~\cite{bai2025qwen3}, GLM-4.5V~\cite{hong2025glm}, and Kimi-K2.5~\cite{team2026kimi} show strong performance on standard benchmarks, while proprietary models, such as GPT-5~\cite{singh2025openai}, Claude-Opus-4.6~\cite{anthropic2026opus}, and Gemini-2.5~\cite{comanici2025gemini}, often excel in more challenging tasks, like image editing and cross-modal reasoning. Despite their performance, the security issues of MLLMs are equally critical and remain largely unresolved. The closed-source MLLMs conceal internal mechanisms and vulnerabilities, making it challenging to evaluate their robustness under black-box scenarios.
\vspace{-5pt}
\subsection{Transfer-based Adversarial Attacks on MLLMs}

Black-box adversarial attacks on MLLMs are generally divided into query-based methods and transfer-based methods. The latter crafts adversarial examples on open-source surrogate models and transfers them to closed-source victims without requiring target-side gradients or repeated API queries. AttackVLM \cite{zhao2023evaluating} first demonstrated that image-to-image feature alignment on surrogate vision encoders such as CLIP \cite{radford2021learning} and BLIP \cite{li2022blip} transfers effectively to downstream MLLMs including MiniGPT \cite{zhu2024minigpt} and LLaVA \cite{liu2023visual}. Subsequent works mainly improved transferability from three perspectives. First, several methods enhanced optimization robustness through loss design and augmentation strategies, including sharpness-aware optimization in CWA \cite{chen2022bootstrap}, spectrum-guided augmentation in SSA-CWA \cite{dong2023robust}, large-scale self-supervised alignment in AnyAttack \cite{zhang2025anyattack}, and diffusion-based feature matching in AdvDiffVLM \cite{guo2024efficient}. Second, recent works explored fine-grained local feature alignment to better capture transferable visual semantics. M-Attack \cite{li2025frustratingly} introduced multi-scale local feature matching via random cropping and resizing, while FOA-Attack \cite{jia2025adversarial} formulated local alignment through optimal transport. M-Attack-V2 \cite{zhao2026pushing} further improved transferability using stronger CLIP ensembles and target-aware semantic guidance, representing one of the strongest existing transfer baselines against closed-source MLLMs. Beyond CLIP-style surrogates, recent studies have also investigated dynamic vision-language alignment~\cite{gu2025improving} and broader transfer protocols across black-box VLMs~\cite{hu2025transferable}.

\vspace{-8pt}
\subsection{Frequency-Domain Adversarial Attacks}
\vspace{-5pt}
A parallel line of work explores the frequency domain to improve adversarial transferability. Early studies show that constraining perturbations to low-frequency components alone already yields stronger black-box transfer~\cite{guo2018low}. Spectrum-based input augmentations such as SSA~\cite{dong2023robust}, frequency-domain model augmentation~\cite{long2022frequency}, variance tuning~\cite{wang2021enhancing}, and spatial-momentum schemes~\cite{wang2022enhancing} diversify perturbations across DCT bands, while contrastive and centralized formulations align frequency content of clean and adversarial features for transferable attacks on classifiers~\cite{yang2024faclattack,wu2024towards}. More recent attacks operate directly on the gradient or feature spectrum: TESSER~\cite{guesmi2025tesser} and high-frequency augmentation with hierarchical-gradient fusion~\cite{zheng2025boosting} regularize ViT gradients in spectral subspaces, commonality-oriented gradient optimization~\cite{gao2025boosting} and frequency-guided adaptive gradient attacks~\cite{fu2025frequency} reweight gradient components by cross-model consistency, and lightweight generators inject frequency-domain priors via attention~\cite{11346045}. These methods, however, are designed for CNN/ViT classification or detection tasks and do not address token-level feature alignment in MLLM vision encoders, where redundancy in spatial patch embeddings and surrogate-specific gradient noise jointly limit transfer to closed-source models.

\vspace{-10pt}
\section{Conclusion}
\vspace{-5pt}
We presented FRA-Attack, a transfer-based targeted attack against MLLMs that addresses two distinct sources of overfitting in the same frequency-domain framework. The redundant low-frequency content in spatial patch embeddings is bypassed by aligning only the high-frequency DCT components through optimal transport, and the surrogate-specific noise concentrated in the high-frequency band of the input gradient is suppressed by a continuous radial decay applied directly to the gradient spectrum. The two components act on different tensors and slot into existing transfer attack pipelines without modifying the surrogate architecture. On a panel of $15$ victim MLLMs covering closed-source standard, closed-source thinking, and open-source families, FRA-Attack achieves the strongest overall performance, and the design choice of a continuous radial decay over per-band statistical clipping is supported by the FGR ablation.



\bibliographystyle{plain}
\bibliography{ref}


\appendix

\section{DCT High-Frequency Energy Visualization}\label{app:fgr-dct-viz}

\begin{figure}[t!]
\centering
\setlength{\tabcolsep}{2pt}
\renewcommand{\arraystretch}{0.4}
\begin{tabular}{@{}c c c c@{}}
& \textbf{Source} & \textbf{Target} & \textbf{Adversarial} \\[1pt]
\rotatebox{90}{\small Big Ben} &
\includegraphics[width=0.2\linewidth]{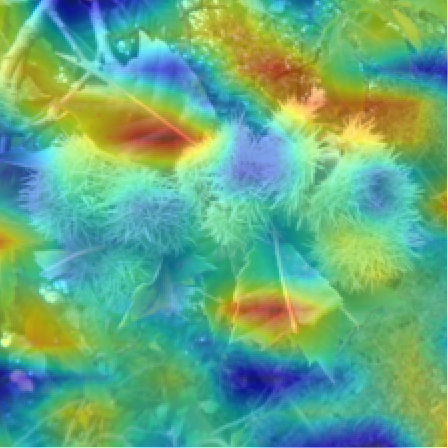} &
\includegraphics[width=0.2\linewidth]{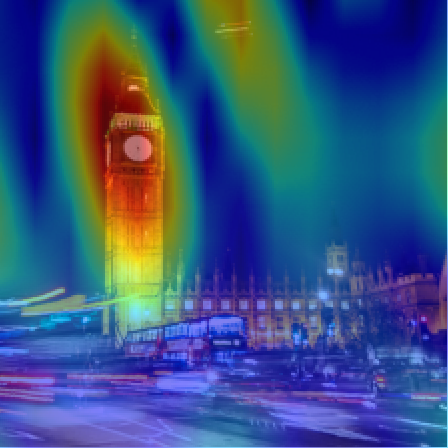} &
\includegraphics[width=0.2\linewidth]{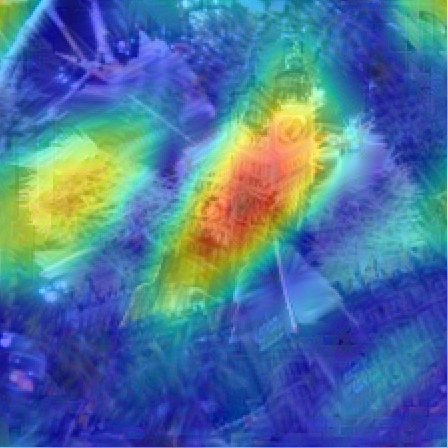}    \\
\rotatebox{90}{\small Hydrant} &
\includegraphics[width=0.2\linewidth]{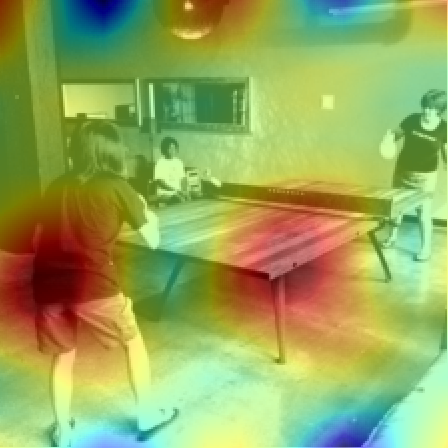} &
\includegraphics[width=0.2\linewidth]{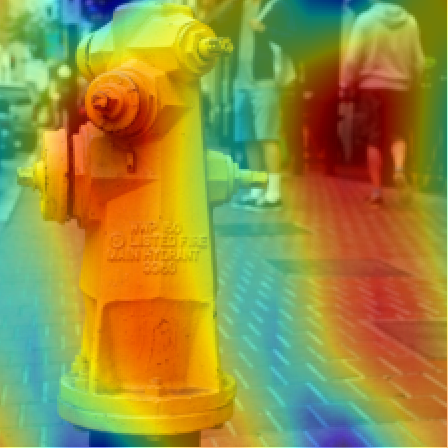} &
\includegraphics[width=0.2\linewidth]{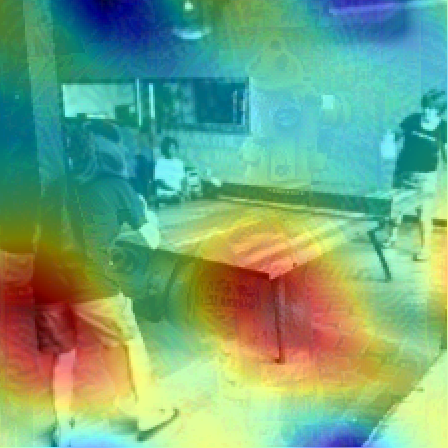}    \\
\rotatebox{90}{\small Kites} &
\includegraphics[width=0.2\linewidth]{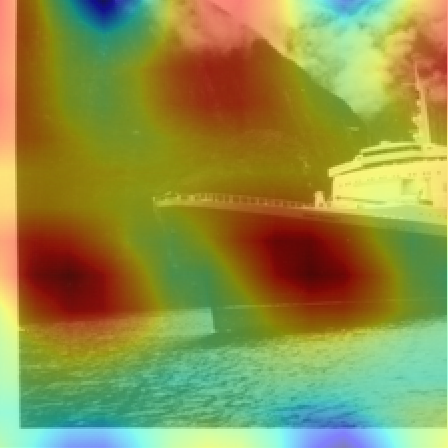} &
\includegraphics[width=0.2\linewidth]{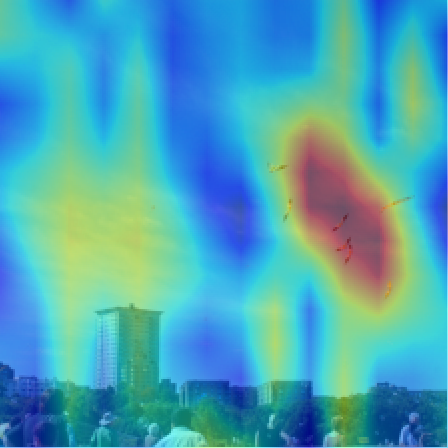} &
\includegraphics[width=0.2\linewidth]{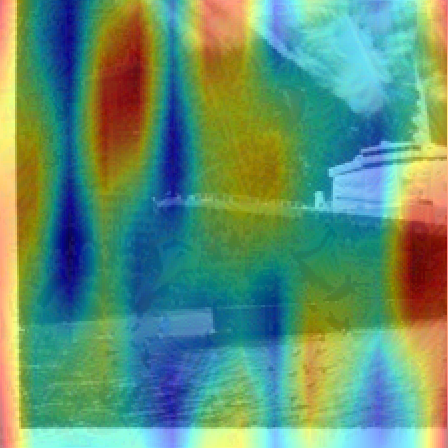}    \\
\rotatebox{90}{\small Skier} &
\includegraphics[width=0.2\linewidth]{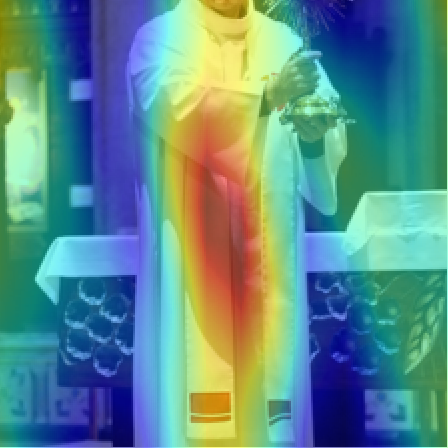} &
\includegraphics[width=0.2\linewidth]{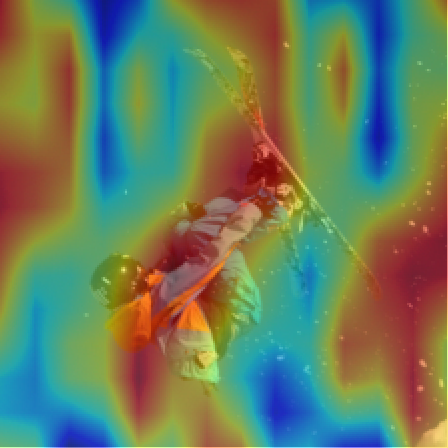} &
\includegraphics[width=0.2\linewidth]{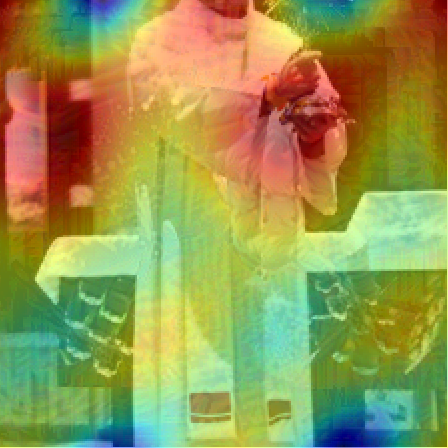}    \\
\rotatebox{90}{\small Atrium} &
\includegraphics[width=0.2\linewidth]{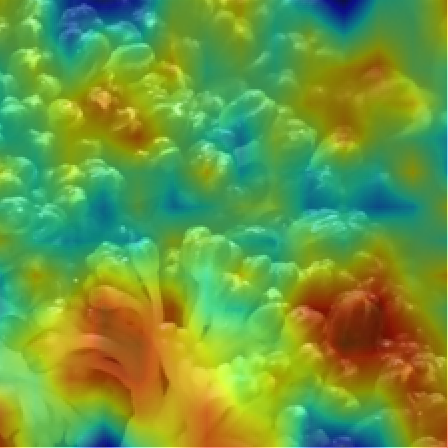} &
\includegraphics[width=0.2\linewidth]{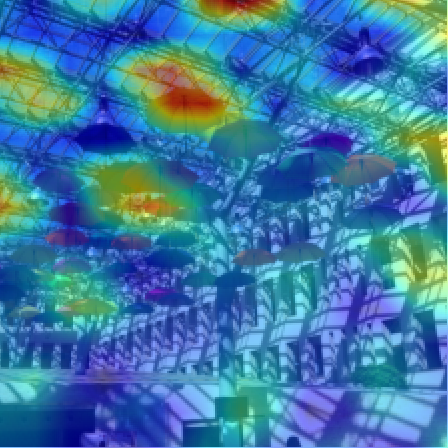} &
\includegraphics[width=0.2\linewidth]{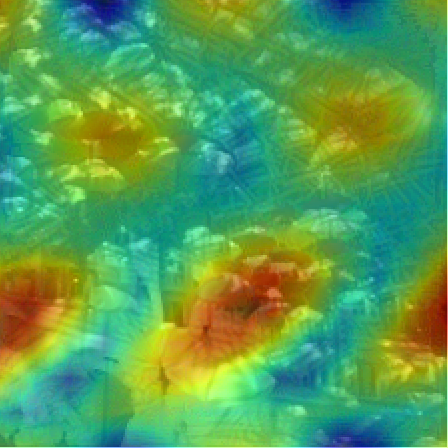}    \\
\rotatebox{90}{\small Mountain road} &
\includegraphics[width=0.2\linewidth]{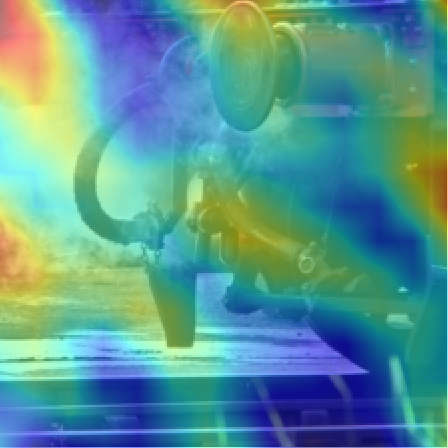} &
\includegraphics[width=0.2\linewidth]{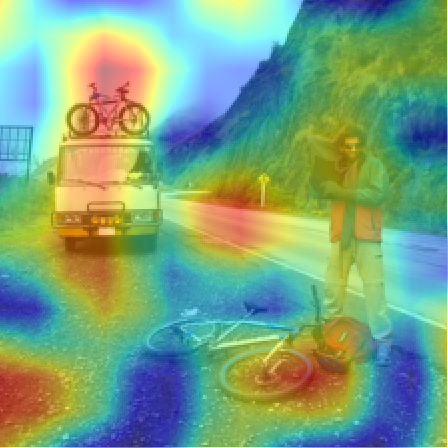} &
\includegraphics[width=0.2\linewidth]{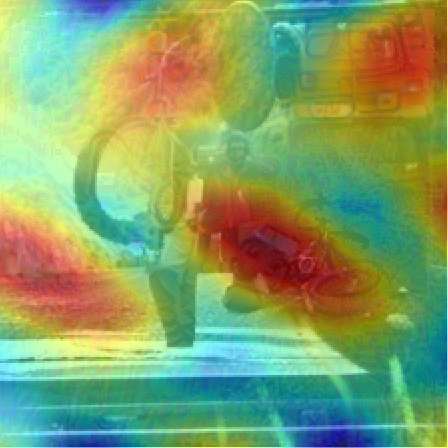}    \\
\end{tabular}
\caption{Per-patch high-frequency DCT energy ($n{=}5$, $\theta{=}20$) of the source, target, and FRA-Attack adversarial example, overlaid on the original image with the \texttt{jet} colormap.}
\label{fig:dct-energy}
\end{figure}

We visualize on real source--target--adversarial triples how the DCT high-frequency selection used by FRA-Attack (Section~\ref{sec:freq}) localizes the salient regions that drive the alignment objective. For each pair we extract the patch embeddings $\bm{E}$ from the ViT-B/16 surrogate, apply the token-wise Type-II DCT (Eq.~\ref{eq:dct}), select the top-$n{=}5$ high-frequency components above the threshold $\theta{=}20$, reconstruct the per-patch energy through a single inverse DCT, and overlay the resulting $14{\times}14$ heatmap on the original $224{\times}224$ image with $50\%$ transparency and the \texttt{jet} colormap. Figure~\ref{fig:dct-energy} reports six representative pairs drawn from the $1{,}000$-pair benchmark. The target column shows that on natural images the high-frequency energy concentrates on the semantic main subject (clock tower, fire hydrant, kite, skier, umbrella array, and roadside vehicles), confirming that the high-frequency DCT band of the patch sequence is a faithful proxy for the perceptually salient region. The source column displays the unrelated source image and serves as a reference. The adversarial column shows that after the FRA-Attack optimization the high-frequency energy shifts away from the source main subject towards the spatial layout of the target, consistent with the cross-victim ASR gain reported in the main tables.

\section{Limitations \& Broader impact}
\paragraph{Limitations.}
Three limitations are worth stating explicitly. First, the surrogate ensemble used in this work consists of three CLIP variants, and our gains depend on the cross-model phase diversity that this ensemble provides; FGR alone on a single CLIP surrogate does not yield the same improvement, so the method is best understood as an ensemble-side regularizer rather than a single-model technique. Second, both proposed components operate on continuous-tone natural images and inherit the assumption that the gradient and the patch embedding spectra are well approximated by smooth radial profiles; the method is not designed for highly structured inputs such as text-rich screenshots or line drawings, where the spectral assumption may not hold. Third, our evaluation focuses on the captioning protocol with GPTScore and KMR judgments. Whether the improvement persists under instruction-grounded protocols such as visual question answering or multi-turn reasoning is left to future work.

\paragraph{Broader impact.}
FRA-Attack improves the transferability of targeted adversarial perturbations against deployed MLLMs, which strengthens the case that current vision-language interfaces should not be relied upon as a sole verification step for safety-critical decisions. We release the method together with the evaluation panel to support the development of stronger detection and certified-defense baselines, rather than as an offensive tool against any specific deployed system.
\section{Detailed Description of FRA-Attack}\label{app:method-details}

\subsection{Algorithm Pseudocode (Extended)}\label{app:algorithm}
Algorithm~\ref{alg:fra} summarizes FRA-Attack at the loop level. We restate it here with full hyperparameter notation and per-step comments to facilitate re-implementation. The complete code, configuration files, and reproduction commands are released in the anonymous repository (Appendix~\ref{app:repro}).

\begin{algorithm}[t]
\caption{FRA-Attack: Frequency-domain Regularized Alignment Attack}
\label{alg:fra}
\begin{algorithmic}[1]
\REQUIRE Source image $\bm{x}_s$, target image $\bm{x}_t$, surrogate ensemble $\{f_j\}_{j=1}^{J}$, perturbation budget $\epsilon$, step size $\alpha$, iterations $N$, frequency threshold $\theta$, number of selected frequencies $n$, global weight $w_g$, local weight $w_l$, FGR decay exponent $p$, momentum decay $\mu$, dynamic-weighting temperature $T$
\ENSURE Adversarial image $\bm{x}_{adv}$

\STATE $\bm{\delta} \leftarrow \bm{0}$,\; $\bm{m} \leftarrow \bm{0}$

\FOR{$t = 1, \ldots, N$}
    \FOR{$j = 1, \ldots, J$}
        \STATE \COMMENT{Extract global and local features}
        \STATE $\bm{g}_j^{(s)}, \bm{E}_j^{(s)} \leftarrow f_j(\bm{x}_s + \bm{\delta})$;\quad $\bm{g}_j^{(t)}, \bm{E}_j^{(t)} \leftarrow f_j(\bm{x}_t)$

        \STATE \COMMENT{DCT and high-frequency selection (Eqs.~\ref{eq:dct}--\ref{eq:local_feat})}
        \STATE $\bm{L}_j^{(s)} \leftarrow \operatorname{HighFreqDCT}(\bm{E}_j^{(s)},\; \theta,\; n)$;\quad $\bm{L}_j^{(t)} \leftarrow \operatorname{HighFreqDCT}(\bm{E}_j^{(t)},\; \theta,\; n)$

        \STATE \COMMENT{Per-model loss}
        \STATE $\mathcal{L}_j^{global} \leftarrow 1 - \cos(\bm{g}_j^{(s)}, \bm{g}_j^{(t)})$
        \STATE $\mathcal{L}_j^{freq} \leftarrow \operatorname{OT}(\bm{L}_j^{(s)}, \bm{L}_j^{(t)})$;\quad see Eq.~\eqref{eq:lfreq}
        \STATE $\mathcal{L}_j \leftarrow w_g \cdot \mathcal{L}_j^{global} + w_l \cdot \mathcal{L}_j^{freq}$
    \ENDFOR

    \STATE \COMMENT{Dynamic ensemble weighting~\cite{jia2025adversarial}}
    \STATE $S_j^{(t)} \leftarrow \mathcal{L}_j^{(t)} / \mathcal{L}_j^{(t-1)}, \;\forall j$
    \STATE $W_j \leftarrow J \cdot \operatorname{Softmax}\!\left(\{S_j^{(t)} / T\}_{j=1}^{J}\right)_j, \;\forall j$;\quad $\mathcal{L} \leftarrow \sum_{j=1}^{J} W_j \cdot \mathcal{L}_j$

    \STATE \COMMENT{Frequency-domain gradient regularization (\S\ref{sec:fgr})}
    \STATE $\tilde{\bm{\nabla}} \leftarrow \operatorname{FGR}(\nabla_{\bm{\delta}} \mathcal{L};\; p)$

    \STATE \COMMENT{MI-FGSM update with regularized gradient}
    \STATE $\bm{m} \leftarrow \mu \cdot \bm{m} + \tilde{\bm{\nabla}} / \|\tilde{\bm{\nabla}}\|_1$
    \STATE $\bm{\delta} \leftarrow \operatorname{clip}\!\left(\bm{\delta} - \alpha \cdot \operatorname{sign}(\bm{m}),\; -\epsilon,\; +\epsilon\right)$
\ENDFOR

\STATE \RETURN $\operatorname{clamp}(\bm{x}_s + \bm{\delta},\; 0,\; 1)$
\end{algorithmic}
\end{algorithm}

\subsection{Surrogate Model Details}\label{app:surrogate-details}
Table~\ref{tab:surrogate-details} reports the three CLIP variants used as the surrogate ensemble, including release source, embedding dimension, and image preprocessing. All three backbones share the same input resolution $224\times 224$, and we apply the same three-stage random crop augmentation across them.

\begin{table}[ht]
\centering
\caption{Surrogate model details. All variants use the same input resolution and augmentation pipeline; only the backbone differs.}
\label{tab:surrogate-details}
\small
\setlength{\tabcolsep}{4pt}
\begin{tabular}{l|c|c|c|c}
\toprule
\textbf{Surrogate} & \textbf{Source} & \textbf{Patch} & \textbf{Embed.\ $d$} & \textbf{\#\,patches $P$} \\
\midrule
CLIP ViT-B/16~\cite{radford2021learning}                  & openai               & 16 & 512  & 196 \\
CLIP ViT-B/32~\cite{radford2021learning}                  & openai               & 32 & 512  & 49  \\
CLIP ViT-g-14-laion2B~\cite{radford2021learning}          & open\_clip / laion2B & 14 & 1024 & 256 \\
\bottomrule
\end{tabular}
\end{table}

\subsection{Full Experimental Setup}\label{app:setup-details}

This subsection collects the implementation details deferred from Section~\ref{sec:settings}.

\paragraph{Source--target pair construction.}
We sample $1{,}000$ source images from the NIPS~2017 Adversarial Attacks and Defenses Competition development set, resize each to $224\times224\times3$, and pair each source with a target image drawn uniformly without replacement from the MSCOCO~\cite{lin2014microsoft} validation split. The same $1{,}000$ pairs are reused across all $7$ baselines, FRA-Attack, and all $15$ victims for strict comparability. To our knowledge, this is the largest victim panel used to date for transfer-based MLLM attacks, jointly covering proprietary, reasoning, and open-weight settings in a single evaluation.

\paragraph{Hyperparameters.}
Perturbation budget $\epsilon = 16/255$ ($\ell_\infty$); step size $\alpha = 1/255$; iterations $N = 300$. The DCT alignment uses high-frequency threshold $\theta = 10$, number of selected high-frequency components $n = 10$, global weight $w_g = 1.0$, local weight $w_l = 0.2$, and Sinkhorn entropic regularization $\lambda = 0.1$. FGR uses the polynomial radial decay (Eq.~\ref{eq:fgr_decay}) with exponent $p = 1.5$. The MI-FGSM momentum decay is $\mu = 1.0$ and the dynamic-weighting temperature is $T = 1.0$. Sweeps over $\theta$, $n$, $w_l$, $p$, $\epsilon$, $N$, and the optimizer are reported in Appendix~\ref{app:ablation-extended}.

\paragraph{M-Attack-V2   configuration.}
We follow the standard targeted transfer protocol of \emph{image-to-image} attack, in which a single source is steered toward a single specified target. The publicly released M-Attack-V2~\cite{zhao2026pushing} deviates from this protocol by enabling target-side retrieval (\texttt{use\_retrieval=True}, \texttt{target\_num=3}): for each target it retrieves $k{-}1$ semantically similar images from an MSCOCO embedding pool and aligns the source against the resulting \emph{image-to-images} set. This both injects external supervision (an auxiliary corpus and a retrieval encoder that lie outside the standard threat model) and changes the optimization objective from matching one target image to matching a target-conditioned cluster, so any ASR difference would conflate the attack itself with the choice of retrieval pool. We therefore set \texttt{use\_retrieval=False} and \texttt{target\_num=1}, keeping every other component (Adam, $\alpha{=}0.005$, $\epsilon{=}16/255$, $N{=}300$, multi-pass $K{=}10$, pooler-weighted alignment with $\beta{=}0.3$, three-stage source crop, target augmentation) at the public defaults. The surrogate ensemble follows Section~\ref{sec:settings}. This configuration is used both in the main result tables and in the component ablation of Section~\ref{sec:ablation_study}.

\paragraph{Hardware and runtime.}
All experiments run on $4\times$ NVIDIA H200 GPUs. A full $1{,}000$-pair attack takes $3$ hours wall-clock under the default settings; victim-side evaluation calls (caption generation and GPTScore matching) are issued through the corresponding vendor APIs.

\section{Extended Ablation Study}\label{app:ablation-extended}

This appendix reports the supplementary ablations referenced from Section~\ref{sec:ablation_study}: (1) attack performance with keyword matching in Figure~\ref{fig:kmr} (Section~\ref{app:ablation-kmr}); (2) the per-cell numbers behind the FGR decay-function and exponent ablation summarized in Figure~\ref{fig:fgr-design} (Section~\ref{app:ablation-fgr-shape}); (3) the DCT alignment hyperparameters $\theta$ and $n$ (Section~\ref{app:ablation-dct-hyper}); (4) the loss balance $w_l$ (Section~\ref{app:ablation-loss-balance}); and (5) the perturbation budget $\epsilon$ and iteration count $N$ (Section~\ref{app:ablation-budget}). Section~\ref{app:ablation-fgr-shape} reports mean ASR on a $6$-victim pilot panel of $100$ source--target pairs (GPT-5.4, GPT-4o, Claude-Opus-4.6, Gemini-2.5-flash, Qwen3-VL-8B, GLM-4.6V) to match the broader victim coverage that motivated the design choice. The remaining sweeps in Sections~\ref{app:ablation-dct-hyper}--\ref{app:ablation-budget} use the same $3$-victim closed-source standard victim group as the main panel (GPT-5.4, Claude-Opus-4.6, Gemini-3-flash), evaluated on the $100$-pair pilot to keep the cluster footprint of the sweeps tractable; the default configuration on this $100$-pair panel matches the $1{,}000$-pair main panel within $\pm 3$ ASR points, so the relative ordering and trends transfer.

\subsection{Attack performance with keyword matching.}
\label{app:ablation-kmr}
\begin{wrapfigure}{r}{0.55\linewidth}
\centering
\vspace{-12pt}
\includegraphics[width=\linewidth]{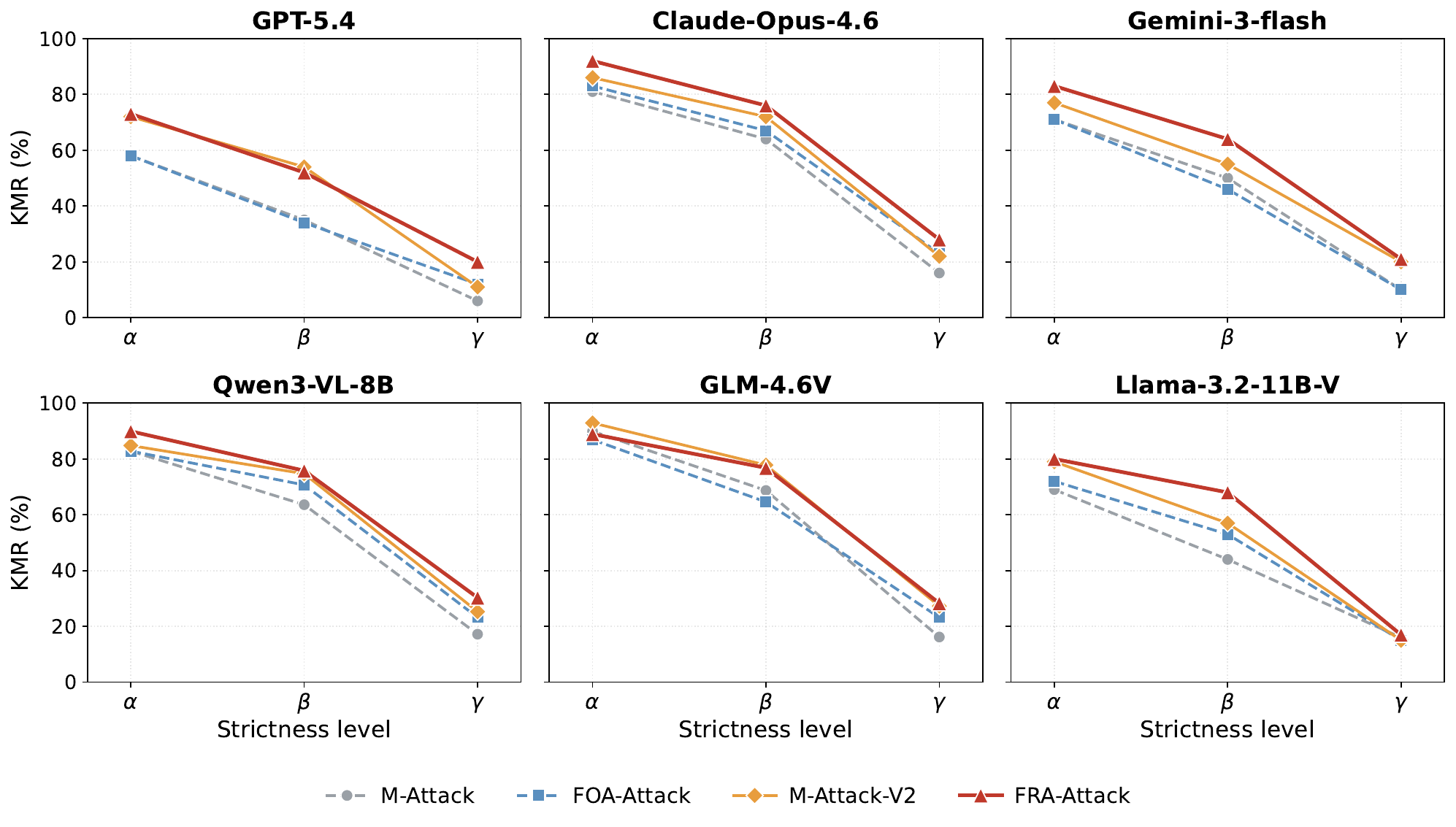}
\vspace{-14pt}
\caption{Three-level keyword matching rate (KMR) on closed-source and open-source MLLMs.}
\label{fig:kmr}
\vspace{-10pt}
\end{wrapfigure}
ASR and AvgSim both depend on a GPT-based caption judge, so we report the keyword matching rate (KMR) as a judgment-free cross-check: KMR counts the fraction of target-image keywords appearing in the victim's response at three strictness levels $\alpha,\beta,\gamma$ (Appendix~\ref{app:prompt}). Figure~\ref{fig:kmr} reports per-victim KMR on the three closed-source standard victims and three open-source models. All six panels share the same shape: every method drops monotonically from $\alpha$ to $\gamma$, FRA-Attack sits at the top of the range, and its gap to M-Attack-V2 widens with strictness, peaking at $\gamma$---the level where the caption must contain every target keyword rather than a subset. This is consistent with the phase-preserving design of FGR (Section~\ref{sec:fgr}), which removes surrogate-specific noise without dampening transferable directions. The three (victim, strictness) combinations where FRA-Attack is not the top method (GPT-5.4 at $\beta$; GLM-4.6V at $\alpha$ and $\beta$) and the full per-cell table appear in Appendix~\ref{app:kmr}.

\begin{table}[ht]
\centering
\caption{Ablation on the FGR decay function and exponent. Mean ASR (\%) over the $6$-victim panel on $100$ source--target pairs (GPT-5.4, GPT-4o, Claude-Opus-4.6, Gemini-2.5-flash, Qwen3-VL-8B, GLM-4.6V).}
\label{tab:ablation-fgr}
\begin{adjustbox}{width=0.92\linewidth}
\small
\begin{tabular}{ll|c|c}
\toprule
\textbf{FGR variant} & \textbf{Hyperparameter} & \textbf{Phase preserving?} & \textbf{Mean ASR (\%)} \\
\midrule
None (no gradient regularization)            & --                            & --   & 62.8 \\
\midrule
Per-band statistical clipping (3 bands)      & $\gamma{=}1.5/1.0/0.5$        & no   & 62.7 \\
Hard sparse threshold (top-$K$\%)            & $K{=}50/30/30$                & no   & 58.2 \\
Reciprocal radial $1/(1+\beta d)$            & $\beta{=}2.0$                 & yes  & 66.8 \\
Sigmoid radial $\sigma(-\beta(d-c))$         & $\beta{=}6,\,c{=}0.5$         & yes  & 65.7 \\
\midrule
Polynomial radial $(1-d)^p$                  & $p{=}0.5$                     & yes  & 60.5 \\
Polynomial radial $(1-d)^p$                  & $p{=}1.0$                     & yes  & 65.0 \\
Polynomial radial $(1-d)^p$                  & $p{=}1.5$ \textbf{(default)}  & yes  & \textbf{69.0} \\
Polynomial radial $(1-d)^p$                  & $p{=}2.0$                     & yes  & 63.0 \\
Polynomial radial $(1-d)^p$                  & $p{=}2.5$                     & yes  & 57.0 \\
Polynomial radial $(1-d)^p$                  & $p{=}3.0$                     & yes  & 53.5 \\
\bottomrule
\end{tabular}
\end{adjustbox}
\end{table}

\subsection{FGR Decay Function and Exponent: Per-cell Numbers}\label{app:ablation-fgr-shape}
Table~\ref{tab:ablation-fgr} provides the per-cell numbers behind Figure~\ref{fig:fgr-design}. The polynomial decay $\phi(d) = (1-d)^p$ used by FRA-Attack is compared against alternative implementations: two non-radial variants illustrate the failure mode of clamp- or threshold-based regularization---(i) per-band statistical clipping that partitions the frequency plane into low/mid/high bands by thresholds $\tau_l < \tau_h$ and clips each band $\mathcal{B}$ within $\mu_{\mathcal{B}} \pm \gamma_{\mathcal{B}} \sigma_{\mathcal{B}}$; (ii) a magnitude-based hard sparse threshold that retains the top $K$\% coefficients in each band and zeros the rest---and two continuous, phase-preserving radial alternatives: (iii) a reciprocal decay $\phi(d) = 1/(1+\beta d)$ and (iv) a sigmoid decay $\phi(d) = \sigma(-\beta(d-c))$. The polynomial decay is additionally swept over $p \in \{0.5, 1.0, 2.0, 2.5, 3.0\}$ around the default $p{=}1.5$.
\noindent The two findings (phase-preserving radial decay is the only family that helps; $p{=}1.5$ is a sharp optimum) are stated and discussed in Section~\ref{sec:ablation_study}.

\subsection{Sensitivity Analysis of DCT Alignment}\label{app:ablation-dct-hyper}

\begin{wraptable}{t}{0.5\linewidth}
\vspace{-15pt}
    \centering
    \caption{Sensitivity Analysis of DCT alignment.}
    \label{tab:ablation-dct}
    \begin{adjustbox}{width=0.8\linewidth}
    \small
    \begin{tabular}{l|cccc}
    \toprule
    \textbf{Sweep} & \multicolumn{4}{c}{\textbf{Value}} \\
    \midrule
    Threshold $\theta$  & 5    & \textbf{10}   & 15   & 20   \\
    Avg ASR (\%)        & 60.3 & \textbf{60.0} & 60.7 & 60.3 \\
    \midrule
    \#\,components $n$  & 5    & \textbf{10}   & 15   & 20   \\
    Avg ASR (\%)        & 58.3 & \textbf{60.0} & 65.0 & 60.0 \\
    \bottomrule
    \end{tabular}
    \end{adjustbox}
    \vspace{-10pt}
\end{wraptable}
Table~\ref{tab:ablation-dct} sweeps the high-frequency threshold $\theta$ and the number of selected high-frequency components $n$. ASR (\%) are averaged across GPT-5.4, Claude-Opus-4.6, and Gemini-3-flash on the $100$-pair pilot panel; the default $\theta{=}n{=}10$ matches the $1{,}000$-pair main panel within $\pm 3$ ASR. The default configuration (in bold) is the same FRA-Attack run reported in the main result tables; sweeps reuse the same source--target pairs and only change the indicated hyperparameter.\par
The DCT alignment is robust to both hyperparameters: across the $\theta$ sweep all four configurations stay within $\pm 0.7$ ASR of the default, indicating that the high-frequency band is well-separated from the low-frequency content and the exact cutoff matters little. The $n$ sweep shows a mild peak at $n{=}15$ ($+5.0$ over the default), suggesting the budget for selected components could be increased; we keep the default $n{=}10$ in the main tables for consistency with the $1{,}000$-pair main panel and to keep the DCT alignment cost in the same range as the $\texttt{[CLS]}$ alignment term.

\subsection{Loss Balance}
\label{app:ablation-loss-balance}

\begin{wraptable}{h}{0.5\linewidth}
\vspace{-15pt}
\centering
\caption{Ablation on the local loss weight $w_l$.}
\label{tab:ablation-loss-balance}
\begin{adjustbox}{width=0.8\linewidth}
\small
\begin{tabular}{l|ccccc}
\toprule
\textbf{Sweep} & \multicolumn{5}{c}{\textbf{Value}} \\
\midrule
Local weight $w_l$  & 0.1  & \textbf{0.2}  & 0.5  & 1.0  & 2.0  \\
Avg ASR (\%)        & 61.3 & \textbf{60.0} & 62.7 & 63.0 & 57.3 \\
\bottomrule
\end{tabular}
\end{adjustbox}
\end{wraptable}
Table~\ref{tab:ablation-loss-balance} sweeps the local loss weight $w_l$ that controls the contribution of the high-frequency DCT alignment term relative to the global \texttt{[CLS]} cosine term (with $w_g = 1.0$ fixed). The default value is $w_l = 0.2$, chosen so that the two loss terms have comparable scale at the start of optimization. ASR (\%) are averaged across GPT-5.4, Claude-Opus-4.6, and Gemini-3-flash on the $100$-pair pilot panel; the default value matches the $1{,}000$-pair main panel within $\pm 3$ ASR.\par
The local loss weight is non-critical inside a wide range $w_l \in [0.1, 1.0]$, where ASR varies by less than $3$ points around the default. Performance only drops noticeably when $w_l$ is pushed to $2.0$, at which point the high-frequency DCT term dominates the global \texttt{[CLS]} alignment and the perturbation drifts away from the coarse target semantics. We therefore keep $w_l = 0.2$, which keeps the two terms balanced at the initialization scale and matches the choice used in the main panel.

\subsection{Perturbation Budget and Iteration Count}\label{app:ablation-budget}

\begin{wraptable}{h}{0.5\linewidth}
\vspace{-15pt}
\centering
\caption{Sensitivity of FRA-Attack to the perturbation budget $\epsilon$ and iteration count $N$. }
\label{tab:ablation-budget}
\begin{adjustbox}{width=0.85\linewidth}
\small
\begin{tabular}{l|cccc}
\toprule
\textbf{Budget $\epsilon$ ($\ell_\infty$)} & $4/255$ & $8/255$ & $16/255$ \textbf{(default)} & $32/255$ \\
\midrule
Avg ASR (\%) & 4.0 & 28.3 & \textbf{60.0} & 80.0 \\
\midrule
\textbf{Iterations $N$} & 50 & 100 & 300 \textbf{(default)} & 500 \\
\midrule
Avg ASR (\%) & 29.0 & 48.7 & \textbf{60.0} & 62.3 \\
\bottomrule
\end{tabular}
\end{adjustbox}
\end{wraptable}
Table~\ref{tab:ablation-budget} reports the sensitivity of FRA-Attack to the perturbation budget $\epsilon$ and the iteration count $N$. The default values are $\epsilon = 16/255$ and $N = 300$.
Average ASR (\%) across GPT-5.4, Claude-Opus-4.6, and Gemini-3-flash on the $100$-pair pilot panel; the default $\epsilon{=}16/255,\,N{=}300$ matches the $1{,}000$-pair main panel within $\pm 3$ ASR.
\noindent FRA-Attack scales monotonically with the perturbation budget: $\epsilon = 4/255$ is too small to escape the natural caption neighbourhood ($4.0$ ASR), $\epsilon = 8/255$ already lands the perturbation in the targeted basin for roughly a quarter of pairs, the default $\epsilon = 16/255$ recovers the main result, and doubling to $\epsilon = 32/255$ adds another $20$ points but is far above the conventional imperceptibility regime. The iteration sweep is more saturating: between $N = 300$ (default) and $N = 500$ ASR moves by only $+2.3$ points, indicating that the attack is close to converged at the default budget and the additional cost of $N = 500$ is not justified.

\section{More Comparison Results under Varied Thresholds}\label{app:thresholds}

The main results in Section~\ref{sec:main_results} use the conventional GPTScore success threshold of $0.5$. We report ASR and AvgSim under three additional thresholds for completeness: a permissive threshold $0.3$ (Tables~\ref{tab:closed_03}--\ref{tab:open_03}), a strict threshold $0.7$ (Tables~\ref{tab:closed_07}--\ref{tab:open_07}), and a very strict threshold $0.9$ (Tables~\ref{tab:closed_09}--\ref{tab:open_09}). The relative ordering across methods is consistent across all four thresholds.

\begin{table}[h]
\centering
\caption{ASR (\%) and AvgSim at threshold $0.3$ on \emph{closed-source standard} MLLMs. Same evaluation set as Table~\ref{tab:closed}.}
\label{tab:closed_03}
\begin{adjustbox}{width=\linewidth}
\scriptsize
\begin{tabular}{l|cc|cc|cc|cc|cc|cc}
\toprule
\multirow{2}{*}{\textbf{Method}}
& \multicolumn{2}{c|}{GPT-5.2} & \multicolumn{2}{c|}{GPT-5.4} & \multicolumn{2}{c|}{Claude-Opus-4.6}
& \multicolumn{2}{c|}{Claude-Sonnet-4.6}
& \multicolumn{2}{c|}{Gemini-3-flash} & \multicolumn{2}{c}{Gemini-2.5-flash} \\
\cmidrule{2-13}
& ASR & AvgSim & ASR & AvgSim & ASR & AvgSim & ASR & AvgSim & ASR & AvgSim & ASR & AvgSim \\
\midrule
AttackVLM$^{\dagger}$~\cite{zhao2023evaluating}   & 8.0 & 0.060 & 4.0 & 0.044 & 6.0 & 0.032 & 6.0 & 0.038 & 3.0 & 0.025 & 3.0 & 0.033 \\
AdvDiffVLM$^{\dagger}$~\cite{guo2024efficient}    & 7.0 & 0.047 & 4.0 & 0.036 & 7.0 & 0.035 & 7.0 & 0.043 & 8.0 & 0.054 & 7.1 & 0.046 \\
SSA-CWA$^{\dagger}$~\cite{dong2023robust}         & 7.0 & 0.049 & 3.0 & 0.033 & 0.0 & 0.022 & 2.0 & 0.022 & 2.0 & 0.028 & 1.0 & 0.029 \\
AnyAttack$^{\dagger}$~\cite{zhang2025anyattack}   & 7.0 & 0.053 & 9.0 & 0.057 & 4.0 & 0.039 & 7.0 & 0.039 & 4.0 & 0.038 & 5.1 & 0.043 \\
\midrule
M-Attack~\cite{li2025frustratingly}                & 51.6 & 0.272 & 55.4 & 0.268 & 80.5 & 0.433 & 77.0 & 0.402 & 67.2 & 0.310 & 80.6 & 0.407 \\
FOA-Attack~\cite{jia2025adversarial}               & 59.5 & 0.312 & 61.7 & 0.300 & 82.7 & 0.455 & 82.2 & 0.444 & 70.9 & 0.341 & 85.0 & 0.439 \\
M-Attack-V2  ~\cite{zhao2026pushing}          & 69.5 & 0.383 & 71.6 & 0.369 & 92.2 & 0.536 & 87.3 & 0.531 & 84.1 & 0.437 & 91.5 & 0.544 \\
\midrule
\textbf{FRA-Attack (Ours)}                         & \textbf{72.3} & \textbf{0.427} & \textbf{75.4} & \textbf{0.411} & \textbf{94.2} & \textbf{0.613} & \textbf{92.1} & \textbf{0.595} & \textbf{84.4} & \textbf{0.456} & \textbf{92.9} & \textbf{0.545} \\
\bottomrule
\end{tabular}
\end{adjustbox}
\end{table}

\begin{table}[h]
\centering
\caption{ASR (\%) and AvgSim at threshold $0.3$ on \emph{closed-source thinking} MLLMs. Same evaluation set as Table~\ref{tab:thinking}.}
\label{tab:thinking_03}
\begin{adjustbox}{width=0.95\linewidth}
\scriptsize
\begin{tabular}{l|cc|cc|cc}
\toprule
\multirow{2}{*}{\textbf{Method}}
& \multicolumn{2}{c|}{GPT-5.4-thinking} & \multicolumn{2}{c|}{Claude-Opus-4.6-thinking}
& \multicolumn{2}{c}{Gemini-3-flash-thinking} \\
\cmidrule{2-7}
& ASR & AvgSim & ASR & AvgSim & ASR & AvgSim \\
\midrule
M-Attack~\cite{li2025frustratingly}             & 50.9 & 0.264 & 77.1 & 0.414 & 61.4 & 0.306 \\
FOA-Attack~\cite{jia2025adversarial}            & 58.5 & 0.301 & 82.6 & 0.453 & 68.1 & 0.348 \\
M-Attack-V2  ~\cite{zhao2026pushing}       & 70.0 & 0.377 & 89.3 & 0.547 & 80.5 & 0.426 \\
\midrule
\textbf{FRA-Attack (Ours)}                      & \textbf{72.9} & \textbf{0.416} & \textbf{92.7} & \textbf{0.614} & \textbf{79.7} & \textbf{0.456} \\
\bottomrule
\end{tabular}
\end{adjustbox}
\end{table}

\begin{table}[h]
\centering
\caption{ASR (\%) and AvgSim at threshold $0.3$ on \emph{open-source} MLLMs. Same evaluation set as Table~\ref{tab:open}.}
\label{tab:open_03}
\begin{adjustbox}{width=\linewidth}
\scriptsize
\begin{tabular}{l|cc|cc|cc|cc|cc|cc}
\toprule
\multirow{2}{*}{\textbf{Method}}
& \multicolumn{2}{c|}{Qwen3-VL-8B} & \multicolumn{2}{c|}{Qwen3-VL-32B} & \multicolumn{2}{c|}{GLM-4.6V}
& \multicolumn{2}{c|}{Llama-3.2-11B-V} & \multicolumn{2}{c|}{Gemma-3-27B-it}
& \multicolumn{2}{c}{Kimi-K2.5} \\
\cmidrule{2-13}
& ASR & AvgSim & ASR & AvgSim & ASR & AvgSim & ASR & AvgSim & ASR & AvgSim & ASR & AvgSim \\
\midrule
AttackVLM$^{\dagger}$~\cite{zhao2023evaluating}   & 11.0 & 0.073 & 10.0 & 0.058 & 7.0 & 0.035 & 5.0 & 0.025 & 54.0 & 0.258 & 5.0 & 0.027 \\
AdvDiffVLM$^{\dagger}$~\cite{guo2024efficient}    & 7.0 & 0.033 & 8.0 & 0.046 & 9.0 & 0.045 & 6.0 & 0.034 & 68.0 & 0.266 & 6.0 & 0.038 \\
SSA-CWA$^{\dagger}$~\cite{dong2023robust}         & 4.0 & 0.024 & 7.0 & 0.043 & 3.0 & 0.033 & 3.0 & 0.025 & 66.0 & 0.272 & 1.0 & 0.017 \\
AnyAttack$^{\dagger}$~\cite{zhang2025anyattack}   & 6.0 & 0.037 & 6.0 & 0.052 & 6.0 & 0.040 & 6.0 & 0.028 & 56.0 & 0.216 & 4.0 & 0.026 \\
\midrule
M-Attack~\cite{li2025frustratingly}                & 85.1 & 0.509 & 83.4 & 0.459 & 88.5 & 0.518 & 65.6 & 0.354 & 72.0 & 0.318 & 69.6 & 0.364 \\
FOA-Attack~\cite{jia2025adversarial}               & 89.2 & 0.547 & 87.9 & 0.491 & 90.5 & 0.554 & 70.1 & 0.380 & 73.5 & 0.339 & 77.4 & 0.408 \\
M-Attack-V2  ~\cite{zhao2026pushing}          & 93.1 & 0.620 & 89.4 & 0.567 & 95.4 & 0.618 & 80.3 & 0.449 & 80.6 & 0.396 & 81.0 & 0.483 \\
\midrule
\textbf{FRA-Attack (Ours)}                         & \textbf{95.6} & \textbf{0.653} & \textbf{94.8} & \textbf{0.617} & \textbf{97.4} & \textbf{0.687} & \textbf{82.2} & \textbf{0.481} & 77.3 & 0.369 & \textbf{90.4} & \textbf{0.568} \\
\bottomrule
\end{tabular}
\end{adjustbox}
\end{table}

\begin{table}[h]
\centering
\caption{ASR (\%) and AvgSim at threshold $0.7$ on \emph{closed-source standard} MLLMs.}
\label{tab:closed_07}
\begin{adjustbox}{width=\linewidth}
\scriptsize
\begin{tabular}{l|cc|cc|cc|cc|cc|cc}
\toprule
\multirow{2}{*}{\textbf{Method}}
& \multicolumn{2}{c|}{GPT-5.2} & \multicolumn{2}{c|}{GPT-5.4} & \multicolumn{2}{c|}{Claude-Opus-4.6}
& \multicolumn{2}{c|}{Claude-Sonnet-4.6}
& \multicolumn{2}{c|}{Gemini-3-flash} & \multicolumn{2}{c}{Gemini-2.5-flash} \\
\cmidrule{2-13}
& ASR & AvgSim & ASR & AvgSim & ASR & AvgSim & ASR & AvgSim & ASR & AvgSim & ASR & AvgSim \\
\midrule
AttackVLM$^{\dagger}$~\cite{zhao2023evaluating}   & 1.0 & 0.060 & 0.0 & 0.044 & 0.0 & 0.032 & 0.0 & 0.038 & 0.0 & 0.025 & 0.0 & 0.033 \\
AdvDiffVLM$^{\dagger}$~\cite{guo2024efficient}    & 0.0 & 0.047 & 0.0 & 0.036 & 0.0 & 0.035 & 0.0 & 0.043 & 0.0 & 0.054 & 0.0 & 0.046 \\
SSA-CWA$^{\dagger}$~\cite{dong2023robust}         & 0.0 & 0.049 & 0.0 & 0.033 & 0.0 & 0.022 & 0.0 & 0.022 & 0.0 & 0.028 & 0.0 & 0.029 \\
AnyAttack$^{\dagger}$~\cite{zhang2025anyattack}   & 0.0 & 0.053 & 0.0 & 0.057 & 0.0 & 0.039 & 0.0 & 0.039 & 0.0 & 0.038 & 0.0 & 0.043 \\
\midrule
M-Attack~\cite{li2025frustratingly}                & 13.5 & 0.272 & 10.4 & 0.268 & 29.1 & 0.433 & 25.9 & 0.402 & 13.8 & 0.310 & 24.2 & 0.407 \\
FOA-Attack~\cite{jia2025adversarial}               & 17.0 & 0.312 & 13.5 & 0.300 & 33.2 & 0.455 & 32.3 & 0.444 & 16.3 & 0.341 & 28.5 & 0.439 \\
M-Attack-V2  ~\cite{zhao2026pushing}          & 24.2 & 0.383 & 21.8 & 0.369 & 45.4 & 0.536 & 47.9 & 0.531 & 28.1 & 0.437 & \textbf{48.6} & 0.544 \\
\midrule
\textbf{FRA-Attack (Ours)}                         & \textbf{33.6} & \textbf{0.427} & \textbf{26.8} & \textbf{0.411} & \textbf{58.5} & \textbf{0.613} & \textbf{57.3} & \textbf{0.595} & \textbf{32.5} & \textbf{0.456} & 45.5 & \textbf{0.545} \\
\bottomrule
\end{tabular}
\end{adjustbox}
\end{table}

\begin{table}[h]
\centering
\caption{ASR (\%) and AvgSim at threshold $0.7$ on \emph{closed-source thinking} MLLMs.}
\label{tab:thinking_07}
\begin{adjustbox}{width=0.95\linewidth}
\scriptsize
\begin{tabular}{l|cc|cc|cc}
\toprule
\multirow{2}{*}{\textbf{Method}}
& \multicolumn{2}{c|}{GPT-5.4-thinking} & \multicolumn{2}{c|}{Claude-Opus-4.6-thinking}
& \multicolumn{2}{c}{Gemini-3-flash-thinking} \\
\cmidrule{2-7}
& ASR & AvgSim & ASR & AvgSim & ASR & AvgSim \\
\midrule
M-Attack~\cite{li2025frustratingly}             & 11.2 & 0.264 & 28.5 & 0.414 & 14.4 & 0.306 \\
FOA-Attack~\cite{jia2025adversarial}            & 14.3 & 0.301 & 32.5 & 0.453 & 19.0 & 0.348 \\
M-Attack-V2  ~\cite{zhao2026pushing}       & 23.1 & 0.377 & 48.8 & 0.547 & 28.1 & 0.426 \\
\midrule
\textbf{FRA-Attack (Ours)}                      & \textbf{28.9} & \textbf{0.416} & \textbf{59.9} & \textbf{0.614} & \textbf{34.8} & \textbf{0.456} \\
\bottomrule
\end{tabular}
\end{adjustbox}
\end{table}

\begin{table}[h]
\centering
\caption{ASR (\%) and AvgSim at threshold $0.7$ on \emph{open-source} MLLMs.}
\label{tab:open_07}
\begin{adjustbox}{width=\linewidth}
\scriptsize
\begin{tabular}{l|cc|cc|cc|cc|cc|cc}
\toprule
\multirow{2}{*}{\textbf{Method}}
& \multicolumn{2}{c|}{Qwen3-VL-8B} & \multicolumn{2}{c|}{Qwen3-VL-32B} & \multicolumn{2}{c|}{GLM-4.6V}
& \multicolumn{2}{c|}{Llama-3.2-11B-V} & \multicolumn{2}{c|}{Gemma-3-27B-it}
& \multicolumn{2}{c}{Kimi-K2.5} \\
\cmidrule{2-13}
& ASR & AvgSim & ASR & AvgSim & ASR & AvgSim & ASR & AvgSim & ASR & AvgSim & ASR & AvgSim \\
\midrule
AttackVLM$^{\dagger}$~\cite{zhao2023evaluating}   & 1.0 & 0.073 & 1.0 & 0.058 & 0.0 & 0.035 & 1.0 & 0.025 & 14.0 & 0.258 & 0.0 & 0.027 \\
AdvDiffVLM$^{\dagger}$~\cite{guo2024efficient}    & 0.0 & 0.033 & 0.0 & 0.046 & 0.0 & 0.045 & 0.0 & 0.034 & 10.0 & 0.266 & 0.0 & 0.038 \\
SSA-CWA$^{\dagger}$~\cite{dong2023robust}         & 0.0 & 0.024 & 0.0 & 0.043 & 0.0 & 0.033 & 0.0 & 0.025 & 12.0 & 0.272 & 0.0 & 0.017 \\
AnyAttack$^{\dagger}$~\cite{zhang2025anyattack}   & 0.0 & 0.037 & 0.0 & 0.052 & 0.0 & 0.040 & 0.0 & 0.028 & 7.0 & 0.216 & 0.0 & 0.026 \\
\midrule
M-Attack~\cite{li2025frustratingly}                & 43.7 & 0.509 & 33.6 & 0.459 & 44.2 & 0.518 & 23.3 & 0.354 & 14.6 & 0.318 & 22.0 & 0.364 \\
FOA-Attack~\cite{jia2025adversarial}               & 47.7 & 0.547 & 38.7 & 0.491 & 50.5 & 0.554 & 24.6 & 0.380 & 16.6 & 0.339 & 27.4 & 0.408 \\
M-Attack-V2  ~\cite{zhao2026pushing}          & 60.9 & 0.620 & 52.2 & 0.567 & 61.2 & 0.618 & 33.2 & 0.449 & \textbf{24.6} & 0.396 & 40.7 & 0.483 \\
\midrule
\textbf{FRA-Attack (Ours)}                         & \textbf{67.0} & \textbf{0.653} & \textbf{60.1} & \textbf{0.617} & \textbf{74.9} & \textbf{0.687} & \textbf{38.0} & \textbf{0.481} & 20.8 & 0.369 & \textbf{51.2} & \textbf{0.568} \\
\bottomrule
\end{tabular}
\end{adjustbox}
\end{table}

\begin{table}[h]
\centering
\caption{ASR (\%) and AvgSim at threshold $0.9$ on \emph{closed-source standard} MLLMs.}
\label{tab:closed_09}
\begin{adjustbox}{width=\linewidth}
\scriptsize
\begin{tabular}{l|cc|cc|cc|cc|cc|cc}
\toprule
\multirow{2}{*}{\textbf{Method}}
& \multicolumn{2}{c|}{GPT-5.2} & \multicolumn{2}{c|}{GPT-5.4} & \multicolumn{2}{c|}{Claude-Opus-4.6}
& \multicolumn{2}{c|}{Claude-Sonnet-4.6}
& \multicolumn{2}{c|}{Gemini-3-flash} & \multicolumn{2}{c}{Gemini-2.5-flash} \\
\cmidrule{2-13}
& ASR & AvgSim & ASR & AvgSim & ASR & AvgSim & ASR & AvgSim & ASR & AvgSim & ASR & AvgSim \\
\midrule
AttackVLM$^{\dagger}$~\cite{zhao2023evaluating}   & 0.0 & 0.060 & 0.0 & 0.044 & 0.0 & 0.032 & 0.0 & 0.038 & 0.0 & 0.025 & 0.0 & 0.033 \\
AdvDiffVLM$^{\dagger}$~\cite{guo2024efficient}    & 0.0 & 0.047 & 0.0 & 0.036 & 0.0 & 0.035 & 0.0 & 0.043 & 0.0 & 0.054 & 0.0 & 0.046 \\
SSA-CWA$^{\dagger}$~\cite{dong2023robust}         & 0.0 & 0.049 & 0.0 & 0.033 & 0.0 & 0.022 & 0.0 & 0.022 & 0.0 & 0.028 & 0.0 & 0.029 \\
AnyAttack$^{\dagger}$~\cite{zhang2025anyattack}   & 0.0 & 0.053 & 0.0 & 0.057 & 0.0 & 0.039 & 0.0 & 0.039 & 0.0 & 0.038 & 0.0 & 0.043 \\
\midrule
M-Attack~\cite{li2025frustratingly}                & 0.6 & 0.272 & 0.3 & 0.268 & 3.0 & 0.433 & 3.1 & 0.402 & 1.1 & 0.310 & 1.5 & 0.407 \\
FOA-Attack~\cite{jia2025adversarial}               & 0.8 & 0.312 & 0.6 & 0.300 & 4.1 & 0.455 & 3.5 & 0.444 & 1.3 & 0.341 & 2.4 & 0.439 \\
M-Attack-V2  ~\cite{zhao2026pushing}          & 2.2 & 0.383 & 1.5 & 0.369 & 6.5 & 0.536 & 7.0 & 0.531 & 2.3 & 0.437 & \textbf{5.4} & 0.544 \\
\midrule
\textbf{FRA-Attack (Ours)}                         & \textbf{3.3} & \textbf{0.427} & \textbf{3.8} & \textbf{0.411} & \textbf{12.1} & \textbf{0.613} & \textbf{12.1} & \textbf{0.595} & \textbf{4.0} & \textbf{0.456} & 5.2 & \textbf{0.545} \\
\bottomrule
\end{tabular}
\end{adjustbox}
\end{table}

\begin{table}[h]
\centering
\caption{ASR (\%) and AvgSim at threshold $0.9$ on \emph{closed-source thinking} MLLMs.}
\label{tab:thinking_09}
\begin{adjustbox}{width=0.95\linewidth}
\scriptsize
\begin{tabular}{l|cc|cc|cc}
\toprule
\multirow{2}{*}{\textbf{Method}}
& \multicolumn{2}{c|}{GPT-5.4-thinking} & \multicolumn{2}{c|}{Claude-Opus-4.6-thinking}
& \multicolumn{2}{c}{Gemini-3-flash-thinking} \\
\cmidrule{2-7}
& ASR & AvgSim & ASR & AvgSim & ASR & AvgSim \\
\midrule
M-Attack~\cite{li2025frustratingly}             & 0.6 & 0.264 & 2.6 & 0.414 & 1.4 & 0.306 \\
FOA-Attack~\cite{jia2025adversarial}            & 0.8 & 0.301 & 3.9 & 0.453 & 1.8 & 0.348 \\
M-Attack-V2  ~\cite{zhao2026pushing}       & 2.1 & 0.377 & 8.0 & 0.547 & 2.2 & 0.426 \\
\midrule
\textbf{FRA-Attack (Ours)}                      & \textbf{3.3} & \textbf{0.416} & \textbf{15.0} & \textbf{0.614} & \textbf{4.6} & \textbf{0.456} \\
\bottomrule
\end{tabular}
\end{adjustbox}
\end{table}

\begin{table}[h]
\centering
\caption{ASR (\%) and AvgSim at threshold $0.9$ on \emph{open-source} MLLMs.}
\label{tab:open_09}
\begin{adjustbox}{width=\linewidth}
\scriptsize
\begin{tabular}{l|cc|cc|cc|cc|cc|cc}
\toprule
\multirow{2}{*}{\textbf{Method}}
& \multicolumn{2}{c|}{Qwen3-VL-8B} & \multicolumn{2}{c|}{Qwen3-VL-32B} & \multicolumn{2}{c|}{GLM-4.6V}
& \multicolumn{2}{c|}{Llama-3.2-11B-V} & \multicolumn{2}{c|}{Gemma-3-27B-it}
& \multicolumn{2}{c}{Kimi-K2.5} \\
\cmidrule{2-13}
& ASR & AvgSim & ASR & AvgSim & ASR & AvgSim & ASR & AvgSim & ASR & AvgSim & ASR & AvgSim \\
\midrule
AttackVLM$^{\dagger}$~\cite{zhao2023evaluating}   & 0.0 & 0.073 & 0.0 & 0.058 & 0.0 & 0.035 & 0.0 & 0.025 & 10.0 & 0.258 & 0.0 & 0.027 \\
AdvDiffVLM$^{\dagger}$~\cite{guo2024efficient}    & 0.0 & 0.033 & 0.0 & 0.046 & 0.0 & 0.045 & 0.0 & 0.034 & 5.0 & 0.266 & 0.0 & 0.038 \\
SSA-CWA$^{\dagger}$~\cite{dong2023robust}         & 0.0 & 0.024 & 0.0 & 0.043 & 0.0 & 0.033 & 0.0 & 0.025 & 6.0 & 0.272 & 0.0 & 0.017 \\
AnyAttack$^{\dagger}$~\cite{zhang2025anyattack}   & 0.0 & 0.037 & 0.0 & 0.052 & 0.0 & 0.040 & 0.0 & 0.028 & 4.0 & 0.216 & 0.0 & 0.026 \\
\midrule
M-Attack~\cite{li2025frustratingly}                & 7.2 & 0.509 & 2.9 & 0.459 & 4.9 & 0.518 & 2.7 & 0.354 & 7.7 & 0.318 & 1.8 & 0.364 \\
FOA-Attack~\cite{jia2025adversarial}               & 9.4 & 0.547 & 3.6 & 0.491 & 7.0 & 0.554 & 1.9 & 0.380 & 8.4 & 0.339 & 2.8 & 0.408 \\
M-Attack-V2  ~\cite{zhao2026pushing}          & 12.8 & 0.620 & 7.5 & 0.567 & 8.7 & 0.618 & 3.4 & 0.449 & \textbf{10.0} & 0.396 & 6.7 & 0.483 \\
\midrule
\textbf{FRA-Attack (Ours)}                         & \textbf{18.3} & \textbf{0.653} & \textbf{11.7} & \textbf{0.617} & \textbf{17.1} & \textbf{0.687} & \textbf{4.8} & \textbf{0.481} & 9.5 & 0.369 & \textbf{11.0} & \textbf{0.568} \\
\bottomrule
\end{tabular}
\end{adjustbox}
\end{table}

\section{Per-cell Keyword Matching Rate}\label{app:kmr}

This appendix provides the per-cell numbers behind Figure~\ref{fig:kmr} in Section~\ref{sec:main_results}. KMR is reported at three increasingly strict matching levels $\alpha,\beta,\gamma$ defined in Appendix~\ref{app:prompt}; $\alpha$ counts a caption as matched if any target keyword appears, $\beta$ requires a majority, and $\gamma$ requires every target keyword to appear.

\begin{table}[h]
\centering
\caption{Keyword Matching Rate (KMR, \%) on the three closed-source standard victims and three open-source victims at strictness levels $\alpha,\beta,\gamma$. Bold marks the strongest method per cell.}
\label{tab:kmr}
\begin{adjustbox}{width=\linewidth}
\scriptsize
\begin{tabular}{l|ccc|ccc|ccc|ccc|ccc|ccc}
\toprule
\multirow{2}{*}{\textbf{Method}}
& \multicolumn{3}{c|}{GPT-5.4}
& \multicolumn{3}{c|}{Claude-Opus-4.6}
& \multicolumn{3}{c|}{Gemini-3-flash}
& \multicolumn{3}{c|}{Qwen3-VL-8B}
& \multicolumn{3}{c|}{GLM-4.6V}
& \multicolumn{3}{c}{Llama-3.2-11B-V} \\
\cmidrule(lr){2-4}\cmidrule(lr){5-7}\cmidrule(lr){8-10}\cmidrule(lr){11-13}\cmidrule(lr){14-16}\cmidrule(lr){17-19}
& $\alpha$ & $\beta$ & $\gamma$
& $\alpha$ & $\beta$ & $\gamma$
& $\alpha$ & $\beta$ & $\gamma$
& $\alpha$ & $\beta$ & $\gamma$
& $\alpha$ & $\beta$ & $\gamma$
& $\alpha$ & $\beta$ & $\gamma$ \\
\midrule
M-Attack~\cite{li2025frustratingly}                & 58.0 & 35.0 & 6.0  & 81.0 & 64.0 & 16.0 & 71.0 & 50.0 & 10.0 & 82.8 & 63.6 & 17.2 & 89.9 & 68.7 & 16.2 & 69.0 & 44.0 & 16.0 \\
FOA-Attack~\cite{jia2025adversarial}               & 58.0 & 34.0 & 12.0 & 83.0 & 67.0 & 23.0 & 71.0 & 46.0 & 10.0 & 82.8 & 70.7 & 23.2 & 86.9 & 64.6 & 23.2 & 72.0 & 53.0 & 15.0 \\
M-Attack-V2~\cite{zhao2026pushing}                 & 72.0 & \textbf{54.0} & 11.0 & 86.0 & 72.0 & 22.0 & 77.0 & 55.0 & 20.0 & 84.8 & 74.7 & 25.3 & \textbf{92.9} & \textbf{77.8} & 27.3 & 79.0 & 57.0 & 15.0 \\
\midrule
\textbf{FRA-Attack (Ours)}
& \textbf{73.0} & 52.0 & \textbf{20.0}
& \textbf{92.0} & \textbf{76.0} & \textbf{28.0}
& \textbf{83.0} & \textbf{64.0} & \textbf{21.0}
& \textbf{89.9} & \textbf{75.8} & \textbf{30.3}
& 88.9 & 76.8 & \textbf{28.3}
& \textbf{80.0} & \textbf{68.0} & \textbf{17.0} \\
\bottomrule
\end{tabular}
\end{adjustbox}
\end{table}

\section{Detailed Evaluation Prompt}\label{app:prompt}

We adopt the same LLM-as-a-judge evaluation framework as M-Attack~\cite{li2025frustratingly} and FOA-Attack~\cite{jia2025adversarial} for direct comparability. This appendix lists the prompts used for caption generation, GPTScore semantic similarity, and KMR keyword extraction.

\paragraph{Caption-generation prompt.}
We query each victim MLLM with the following single prompt to obtain a free-form description of the input image:
\begin{quote}
\texttt{Describe this image.}
\end{quote}

\paragraph{GPTScore semantic-similarity prompt.}
We use GPT-4o as the judge model. The judge receives the captions of the adversarial example and the target image and outputs a similarity score in $[0, 1]$:
\begin{quote}\small
\texttt{Rate the semantic similarity between the following two texts on a scale from 0 to 1.\\
\textbf{Criteria for similarity measurement:}\\
1.\ Main Subject Consistency: same key subject/object → higher score.\\
2.\ Relevant Description: same context or topic → higher score.\\
3.\ Ignore Fine-Grained Details: do not penalize phrasing or minor variation.\\
4.\ Partial Matches: extra information without contradiction → still high score.\\
5.\ Score Range:\\
\hspace*{1em}1.0: nearly identical in meaning.\\
\hspace*{1em}0.8--0.9: same subject, highly related descriptions.\\
\hspace*{1em}0.7--0.8: same subject, core meaning aligned.\\
\hspace*{1em}0.5--0.7: same subject, different perspectives or missing details.\\
\hspace*{1em}0.3--0.5: related but not highly similar.\\
\hspace*{1em}0.0--0.2: completely different subjects or unrelated.\\
Text 1: \{caption\_adv\}\\
Text 2: \{caption\_tgt\}\\
Output only a single number between 0 and 1. Do not include any explanation or additional text.}
\end{quote}

\paragraph{KMR keyword-extraction prompt.}
We use GPT-4o to extract a small set of content nouns from the target caption, and check whether the same nouns (or their WordNet synsets) appear in the adversarial caption. The extraction prompt is:
\begin{quote}\small
\texttt{From the following description, extract up to three concrete content nouns that identify the main subjects.\\
Output them as a comma-separated list, lowercase, singular form, no extra text.\\
Description: \{caption\_tgt\}}
\end{quote}

\section{Comparison Results on Defense Methods}\label{app:defense}

We evaluate the robustness of FRA-Attack against three representative input-space defenses applied at evaluation time: JPEG compression (quality $75$), Gaussian smoothing (kernel $5\times5$, $\sigma=0.5$), and center cropping (crop ratio $0.9$ followed by upsampling back to $224\times224$). Each defense is applied to the adversarial example before it is fed to the victim MLLM, and the captioning + GPTScore evaluation pipeline of Appendix~\ref{app:prompt} is then run on the defended image. JPEG and Gaussian smoothing act as low-pass filters that suppress high-frequency perturbation energy, while center cropping further perturbs the spatial alignment between the perturbation and the patch grid of the victim ViT. Table~\ref{tab:defense-closed} reports ASR (\%) and AvgSim under each defense for the strongest prior baseline (M-Attack-V2 in the v2 configuration of Section~\ref{sec:settings}) and FRA-Attack on the three closed-source standard MLLMs used in the main evaluation. We focus on closed-source standard models because they are the most realistic deployment target for input-space defenses; extending the evaluation to open-source MLLMs and to learned defenses such as ComDefend is left for future work.

Across all three defenses and all three victim models, FRA-Attack consistently retains higher ASR and AvgSim than M-Attack-V2. The advantage is most pronounced under JPEG compression and center cropping, where FRA-Attack improves ASR@$0.5$ by $+9$ to $+16$ points absolute over V2 (e.g., $+9.3$ on Claude under JPEG, $+16.0$ on Claude under center cropping), supporting our motivation: the FGR component attenuates the high-frequency band of the input gradient and steers the resulting perturbation toward lower-frequency, defense-resistant directions, so a low-pass or geometry-mild defense removes a smaller fraction of the attack energy than for a baseline that allocates more energy to high-frequency components. Gaussian smoothing with a $\sigma=0.5$ kernel is the strongest of the three defenses for both methods, but FRA-Attack still leads on $2$ of $3$ models (Claude $+8.0$, Gemini-3-flash $+9.0$) and is essentially tied on GPT-5.4, where both methods collapse to single-digit ASR.

\begin{table}[h]
\centering
\caption{Attack performance of adversarial images against \emph{closed-source standard} MLLMs after defense processing. ASR (\%) / AvgSim under each defense on $100$ source--target pairs. Defenses: JPEG quality $75$, Gaussian smoothing ($5\times5$, $\sigma=0.5$), and center cropping ratio $0.9$. \textbf{Bold} marks the best ASR per (defense, model) cell.}
\label{tab:defense-closed}
\begin{adjustbox}{width=0.8\linewidth}
\small
\begin{tabular}{l|l|cc|cc|cc}
\toprule
\multirow{2}{*}{\textbf{Defense}} & \multirow{2}{*}{\textbf{Method}}
& \multicolumn{2}{c|}{Claude-Opus-4.6} & \multicolumn{2}{c|}{GPT-5.4} & \multicolumn{2}{c}{Gemini-3-flash} \\
\cmidrule{3-8}
& & ASR & AvgSim & ASR & AvgSim & ASR & AvgSim \\
\midrule
\multirow{2}{*}{JPEG ($q{=}75$)}
  & M-Attack-V2   & 70.7 & 0.570 & 35.0 & 0.339 & 45.0 & 0.420 \\
  & \textbf{FRA-Attack} & \textbf{80.0} & \textbf{0.644} & \textbf{48.0} & \textbf{0.438} & \textbf{55.0} & \textbf{0.478} \\
\midrule
\multirow{2}{*}{Gaussian ($\sigma{=}0.5$)}
  & M-Attack-V2   & 21.0 & 0.215 & \textbf{5.0} & \textbf{0.080} & 13.0 & 0.171 \\
  & \textbf{FRA-Attack} & \textbf{29.0} & \textbf{0.288} & 3.0 & 0.080 & \textbf{22.0} & \textbf{0.214} \\
\midrule
\multirow{2}{*}{Center crop ($r{=}0.9$)}
  & M-Attack-V2   & 67.0 & 0.548 & 47.0 & 0.426 & 51.0 & 0.471 \\
  & \textbf{FRA-Attack} & \textbf{83.0} & \textbf{0.679} & \textbf{59.0} & \textbf{0.497} & \textbf{63.0} & \textbf{0.538} \\
\bottomrule
\end{tabular}
\end{adjustbox}
\end{table}

\section{Commercial MLLM Response Visualizations}\label{app:commercial}
\begin{figure}[t!]\centering
\includegraphics[width=0.8\linewidth]{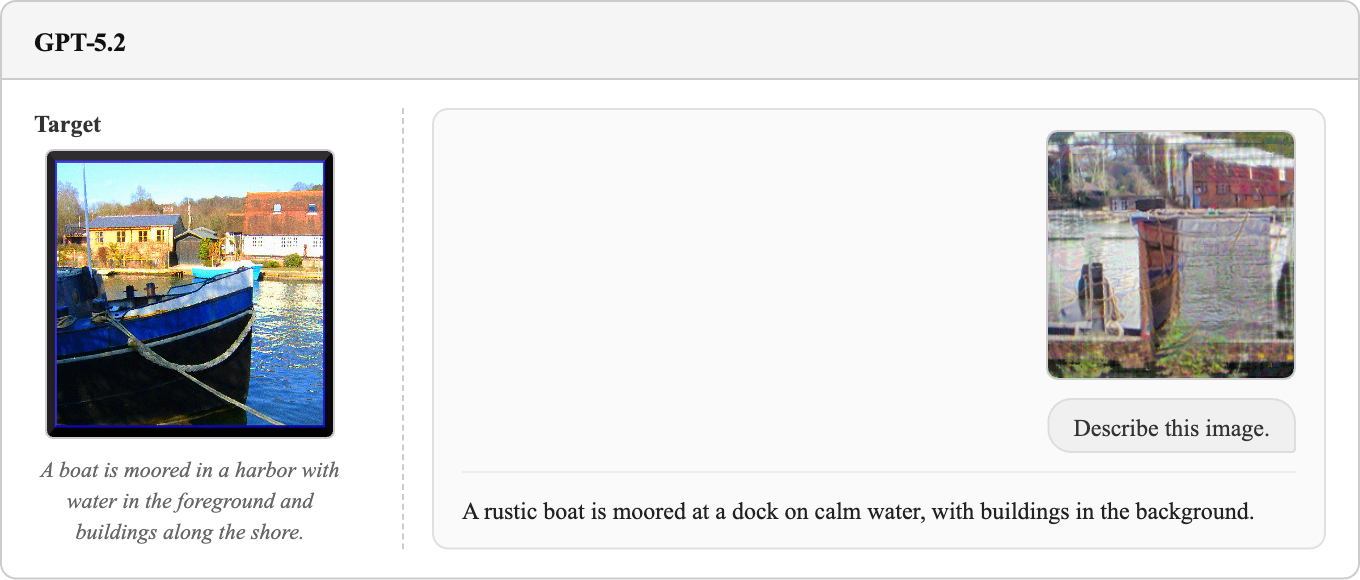}\\[2pt]
\includegraphics[width=0.8\linewidth]{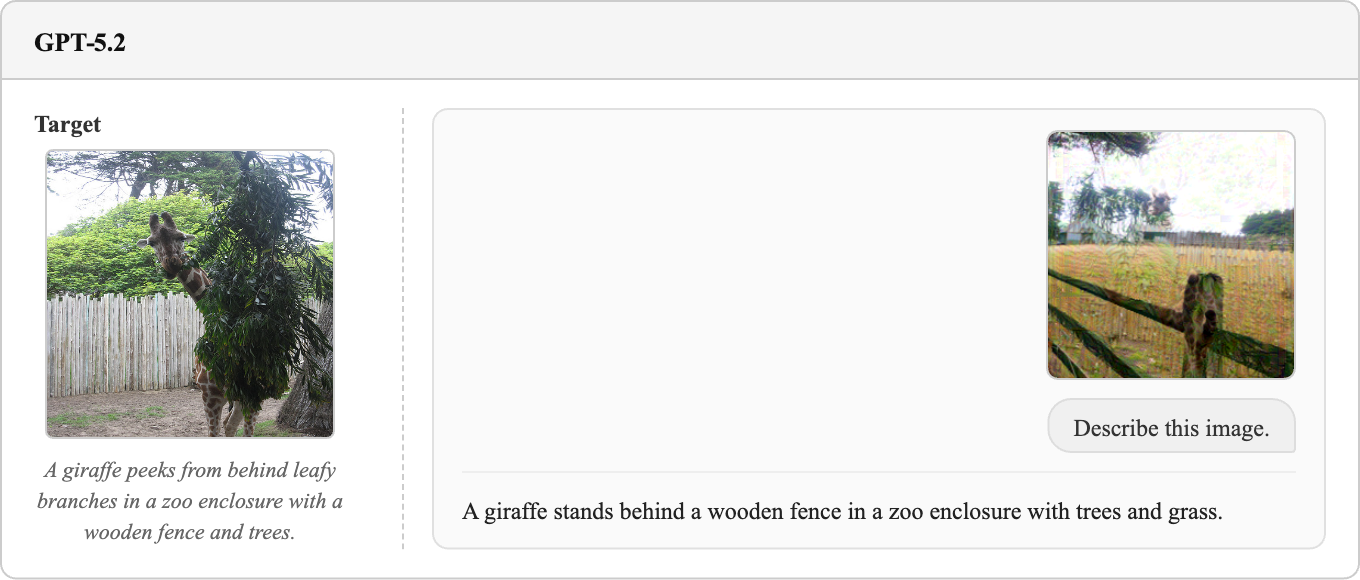}
\caption{GPT-5.2 responses to FRA-Attack adversarial examples.}\label{fig:commercial-gpt52}\end{figure}

\begin{figure}[t!]\centering
\includegraphics[width=0.8\linewidth]{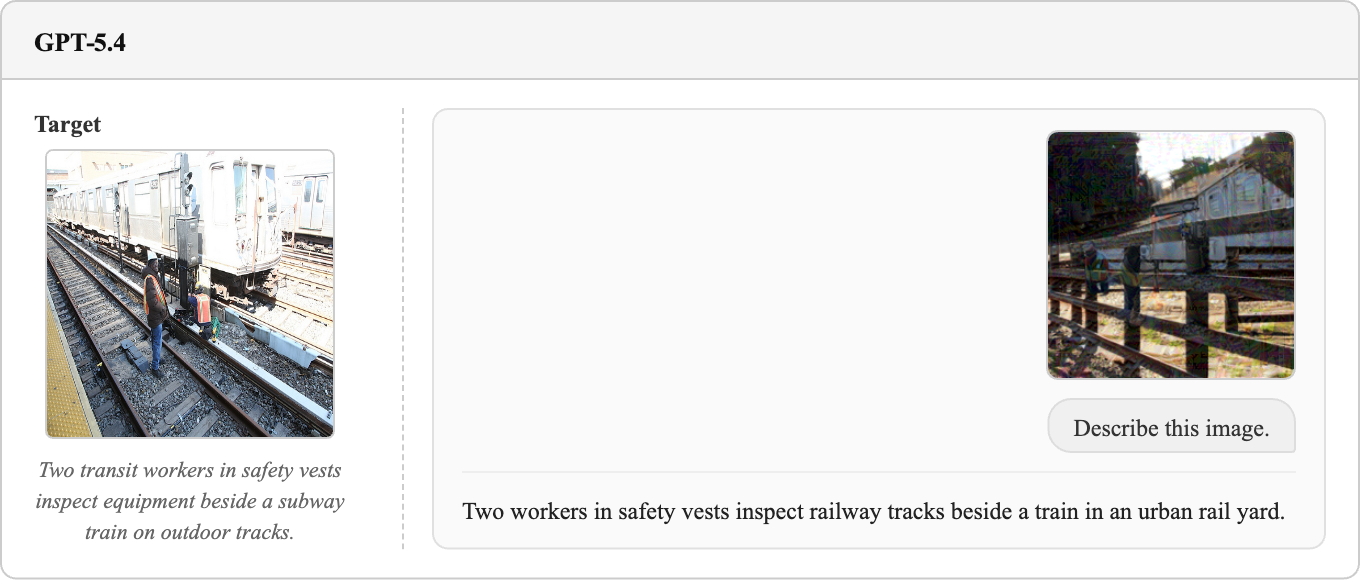}
\caption{GPT-5.4 responses to FRA-Attack adversarial examples.}\label{fig:commercial-gpt54}\end{figure}

\begin{figure}[t!]\centering
\includegraphics[width=0.8\linewidth]{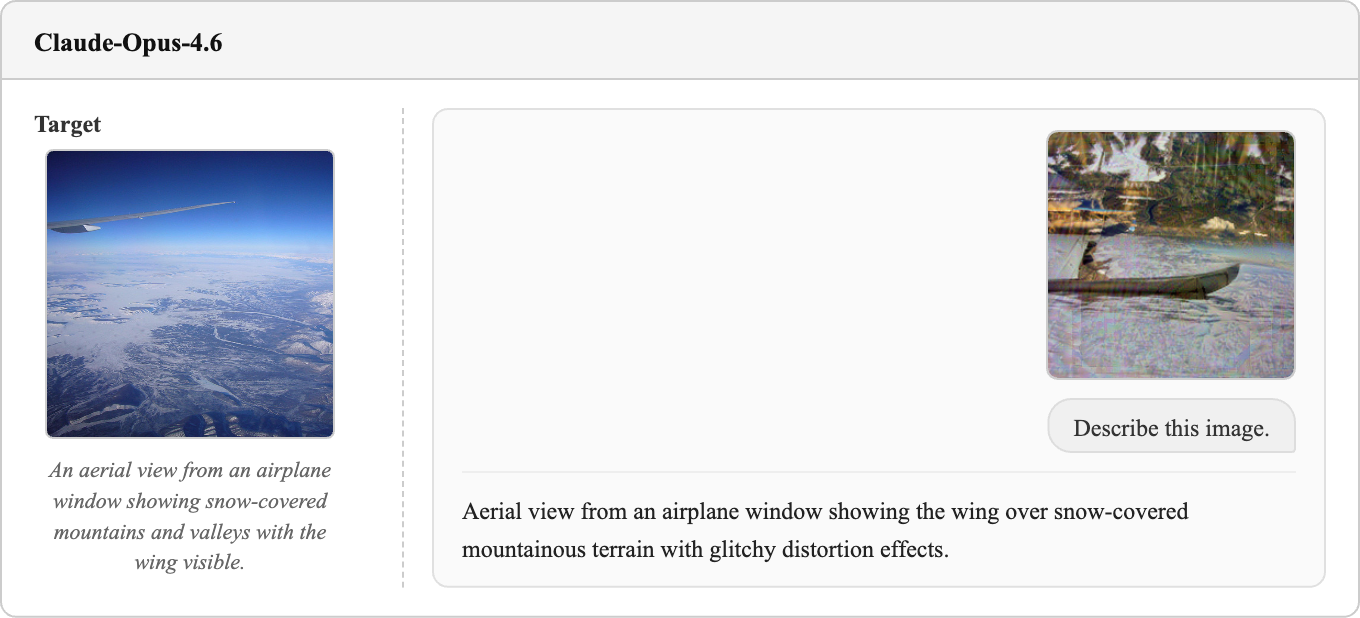}\\[2pt]
\includegraphics[width=0.8\linewidth]{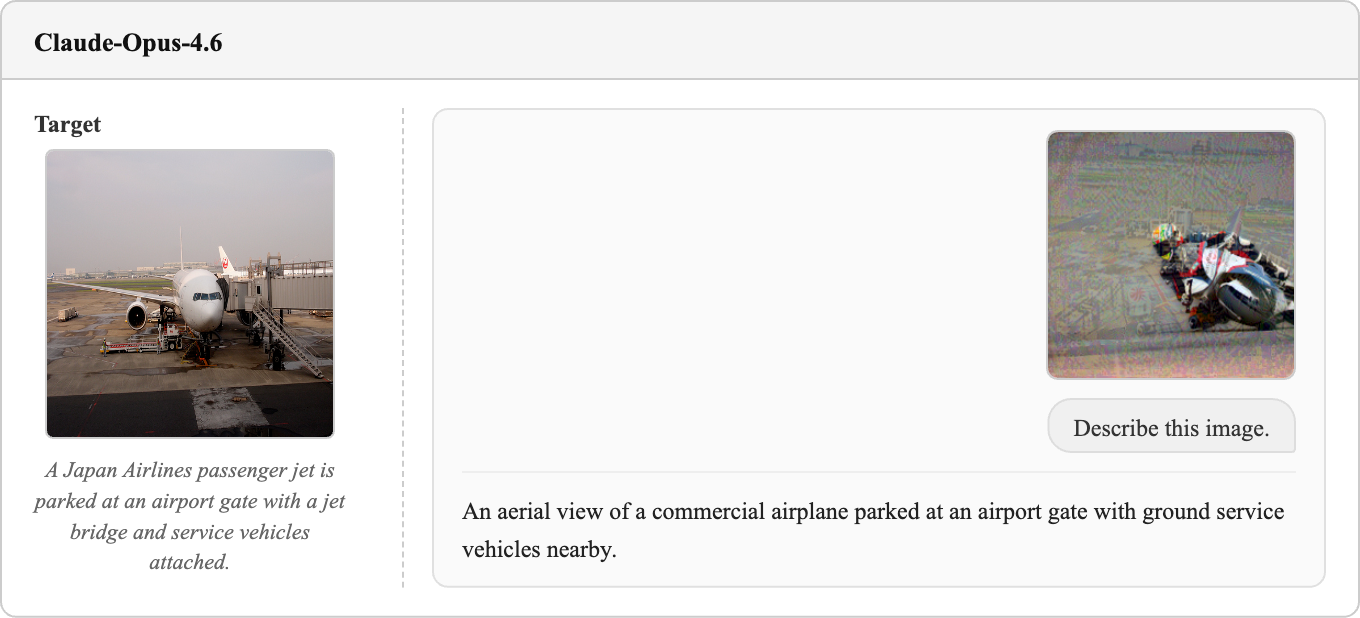}\\[2pt]
\includegraphics[width=0.8\linewidth]{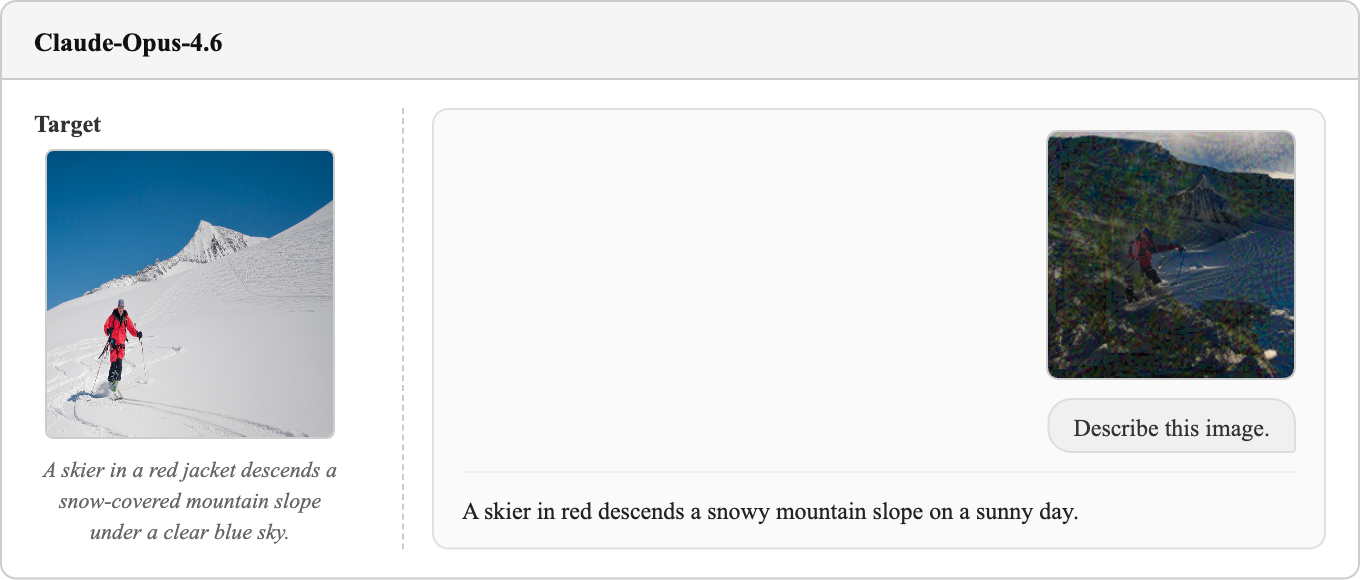}\\[2pt]
\includegraphics[width=0.8\linewidth]{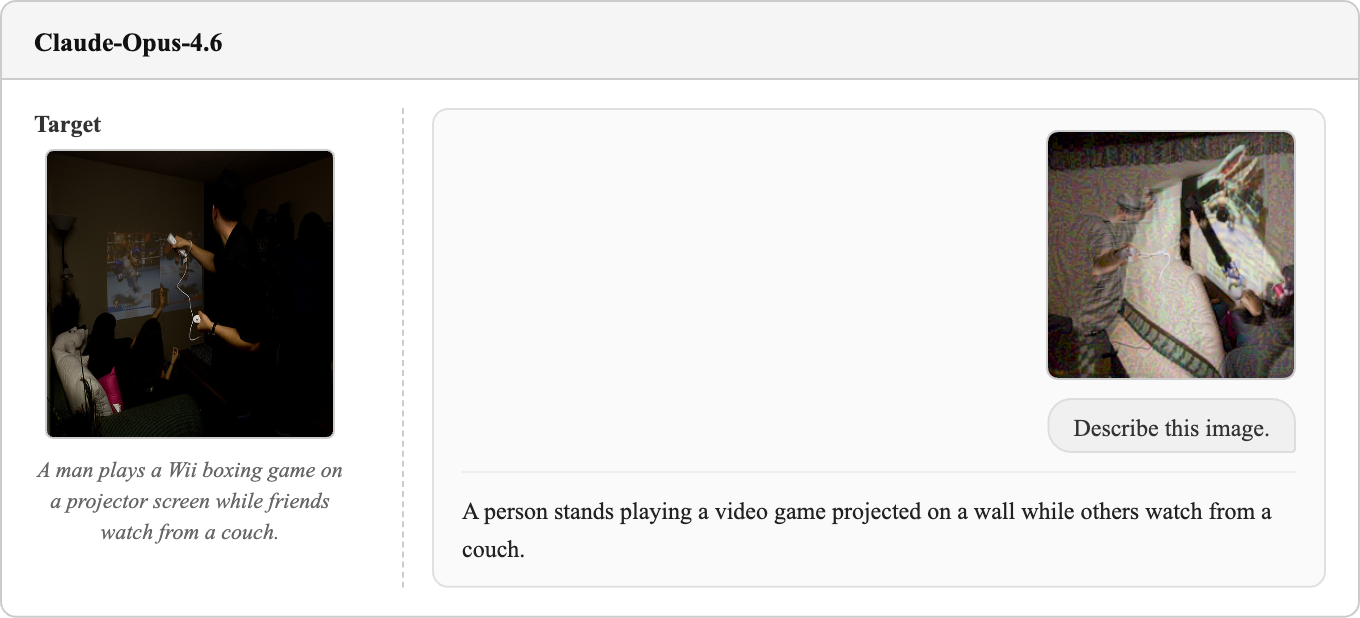}
\caption{Claude-Opus-4.6 responses to FRA-Attack adversarial examples.}\label{fig:commercial-claude-opus}\end{figure}

\begin{figure}[t!]\centering
\includegraphics[width=0.8\linewidth]{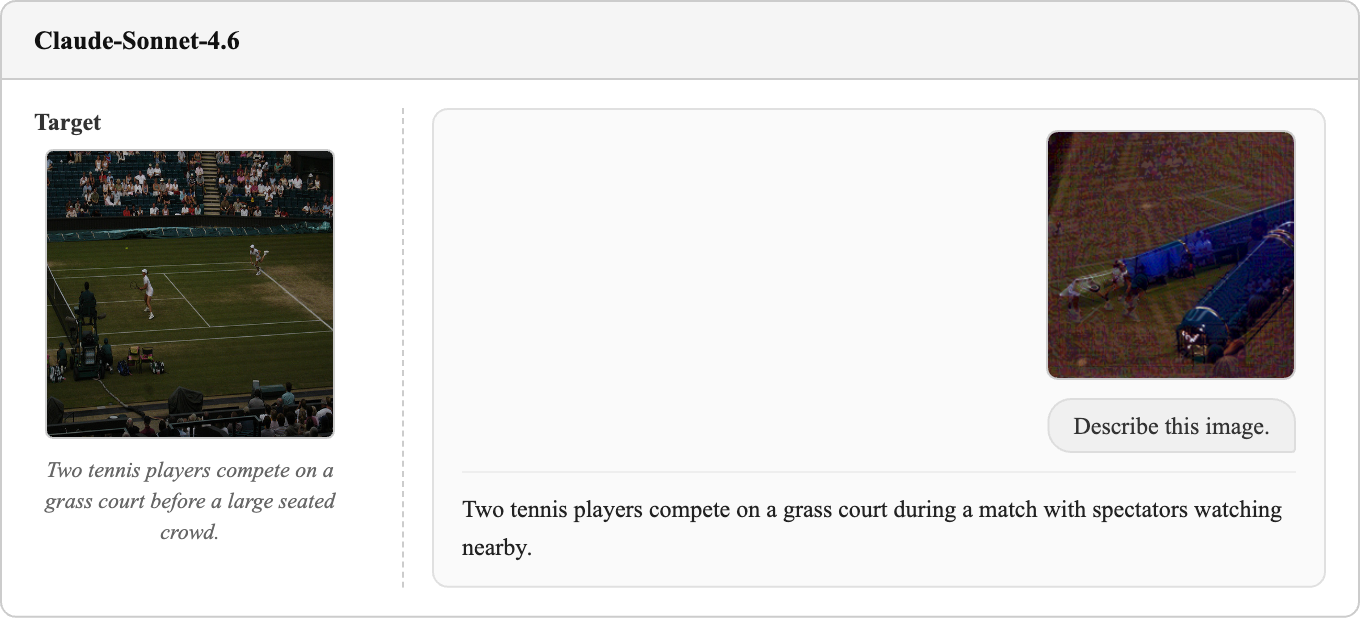}
\caption{Claude-Sonnet-4.6 responses to FRA-Attack adversarial examples.}\label{fig:commercial-claude-sonnet}\end{figure}

\begin{figure}[t!]\centering
\includegraphics[width=0.8\linewidth]{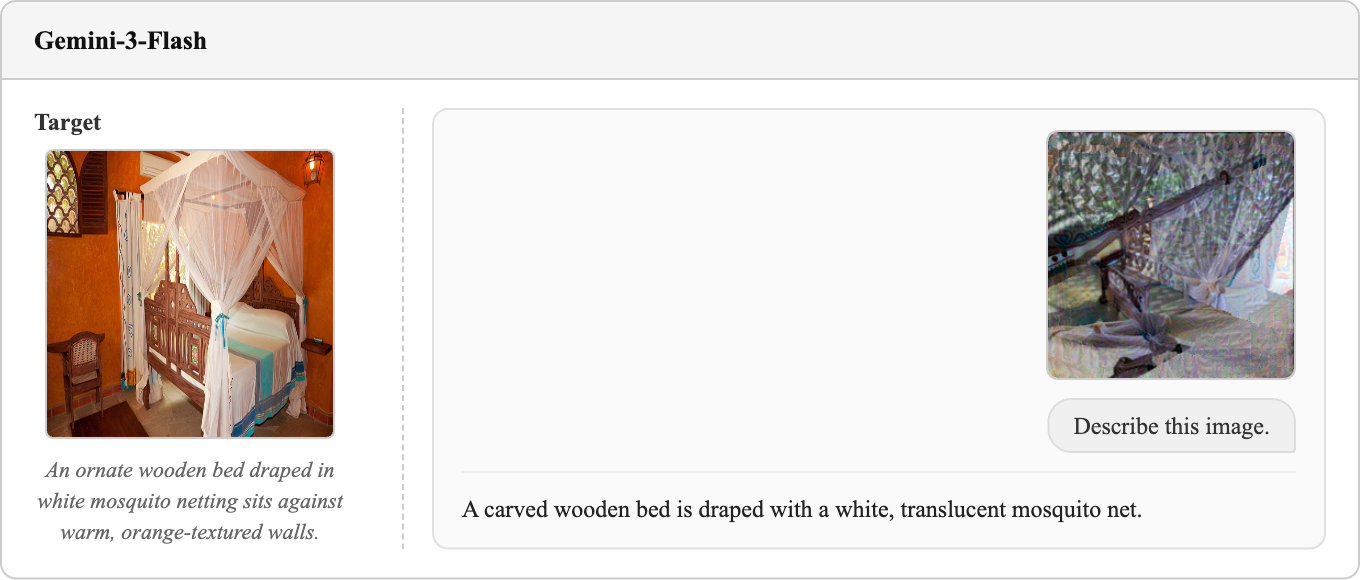}\\[2pt]
\includegraphics[width=0.8\linewidth]{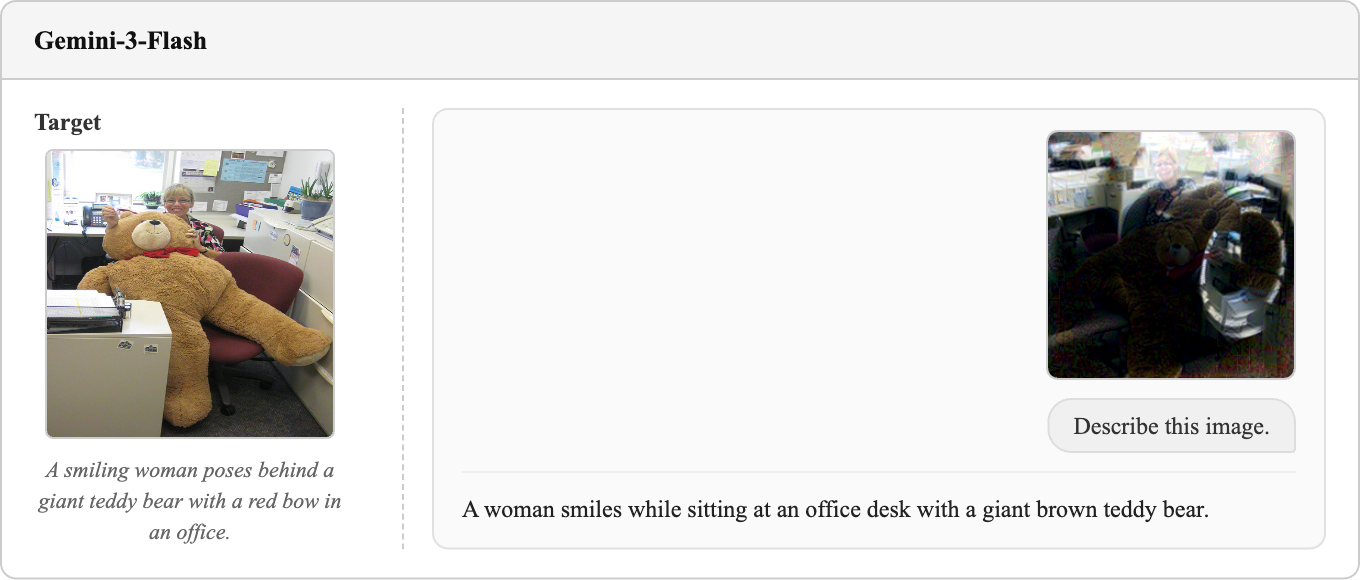}
\caption{Gemini-3-Flash responses to FRA-Attack adversarial examples.}\label{fig:commercial-gemini-flash}\end{figure}

\begin{figure}[t!]\centering
\includegraphics[width=0.8\linewidth]{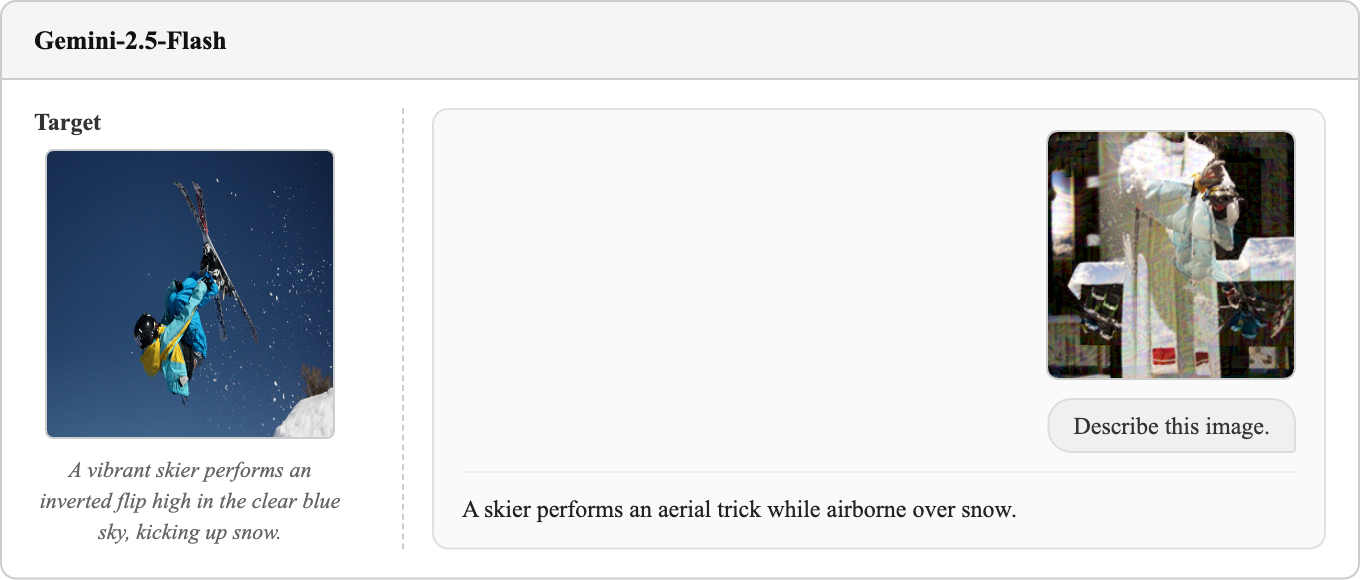}\\[2pt]
\includegraphics[width=0.8\linewidth]{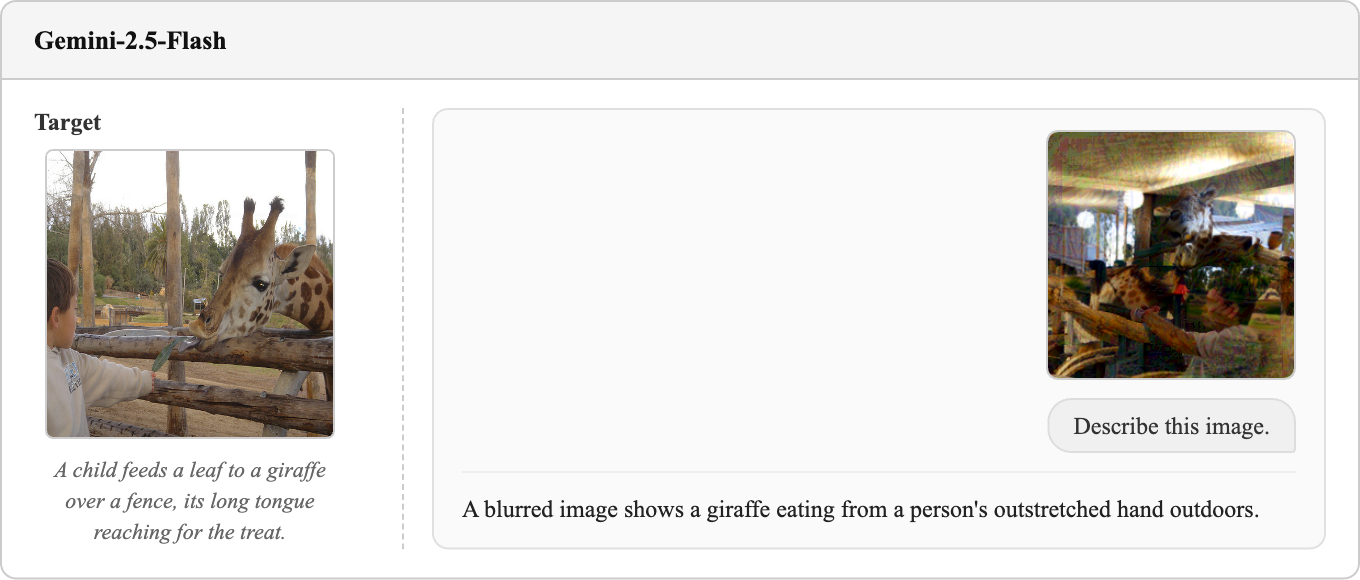}
\caption{Gemini-2.5-Flash responses to FRA-Attack adversarial examples.}\label{fig:commercial-gemini-25-flash}\end{figure}

To complement the quantitative results, we provide rendered conversations with each closed-source MLLM on real source--target pairs from the $1{,}000$-pair benchmark. For every victim, each panel shows the target image, the FRA-Attack adversarial example, and the model's response to the prompt ``Describe this image.''; the responses shown are verbatim outputs collected during the main evaluation.

\section{Theoretical Justification}\label{app:theory}

This appendix gives algebraic justifications for three design choices central to FRA-Attack: (1) why low-frequency DCT components of the patch sequence encode global structure while high-frequency components encode patch-specific fine-grained semantics; (2) why aligning the high-frequency DCT components of patch embeddings is more informative than aligning spatial-domain cluster centers; and (3) why the polynomial radial decay used by FGR preserves the spatial direction of the input gradient, while per-band statistical clipping does not.

\subsection{Low-Frequency DCT Components Encode Global Structure}\label{app:theory-dct}
We justify the claim made in Section~\ref{sec:dct} that low-frequency DCT components of the patch sequence carry global structure shared across patches, while high-frequency components carry patch-specific fine-grained semantics.

\paragraph{Setup.}
Let $\bm{E} = [\bm{e}_0, \ldots, \bm{e}_{P-1}]^\top \in \mathbb{R}^{P \times d}$ denote the patch embeddings of an image, and let $\bm{F}_k \in \mathbb{R}^d$ denote the $k$-th Type-II DCT component along the token dimension defined in Eq.~\ref{eq:dct}. Write the cosine basis as $\psi_k(n) = \cos[\pi(n+\tfrac{1}{2})k/P]$, so that $\bm{F}_k = \sum_{n=0}^{P-1} \psi_k(n)\,\bm{e}_n$.

\paragraph{Claim 1 (DC component equals the global mean).}
The DC coefficient $\bm{F}_0$ is exactly $P$ times the mean patch embedding:
\begin{equation}
  \bm{F}_0 = \sum_{n=0}^{P-1} \psi_0(n)\,\bm{e}_n = \sum_{n=0}^{P-1} \bm{e}_n = P\,\bar{\bm{e}}, \qquad \bar{\bm{e}} = \frac{1}{P}\sum_{n=0}^{P-1} \bm{e}_n.
\end{equation}
$\bm{F}_0$ is therefore the part of the patch sequence that is identical at every position, with no inter-patch variation. For low frequencies $k \approx 0$ the basis $\psi_k(n)$ varies slowly across $n$ and acts as a smoothed running mean, so $\bm{F}_k$ remains dominated by the global content shared across patches.

\paragraph{Claim 2 (high-frequency components are orthogonal to the global mean).}
By the orthogonality of the Type-II DCT basis, for every $k \geq 1$:
\begin{equation}
  \sum_{n=0}^{P-1} \psi_k(n) = 0.
\end{equation}
A patch-uniform signal $\bm{e}_n \equiv \bm{c}$ therefore yields $\bm{F}_k = \bm{c}\sum_n \psi_k(n) = \bm{0}$ for all $k \geq 1$. Equivalently, every $\bm{F}_k$ with $k \geq 1$ is invariant to a constant shift of the patch sequence: replacing $\bm{e}_n$ with $\bm{e}_n + \bm{c}$ leaves $\bm{F}_k$ unchanged. High-frequency components carry, by construction, no information about content shared identically across all patches.

\paragraph{Claim 3 (high-frequency components localize to inter-patch variation).}
The basis $\psi_k(n)$ oscillates at angular frequency $\pi k/P$, and the per-step difference between adjacent samples satisfies
\begin{equation}
  \psi_k(n+1) - \psi_k(n) = -2\sin\!\left[\tfrac{\pi k}{2P}\right]\sin\!\left[\tfrac{\pi k(n+1)}{P}\right],
\end{equation}
whose magnitude grows monotonically with $k$ on $[0, P]$. Consequently $\bm{F}_k$ at large $k$ behaves as a difference operator that responds primarily to inter-patch variation between neighbouring tokens, rather than to the slowly-varying global content captured by low-$k$ components.

\paragraph{Implication for FRA-Attack.}
Claims~1--3 jointly justify the design choice in Section~\ref{sec:freq}: aligning low-frequency DCT components would re-weigh the global mean already covered by the \texttt{[CLS]} cosine, while the top-$n$ high-frequency components carry the orthogonal, patch-wise signal that the global term cannot supply. The empirical confirmation is reported in Appendix~\ref{app:ablation-dct-hyper}: the low/high split threshold $\theta$ is robust within $\pm 0.7$ ASR in the range $[5, 20]$, consistent with a well-separated low/high band predicted by Claims~1 and 2.

\subsection{Polynomial Decay Preserves DCT-Domain Phase Relations}\label{app:theory-fgr}
We show that the continuous radial decay $\phi(d) = (1-d)^p$ used in FGR (Eq.~\ref{eq:fgr_decay}) is equivalent to a separable, smooth filter applied in the spatial domain, and therefore preserves the relative magnitude and sign of every pair of input-gradient coefficients. In contrast, the per-band statistical clipping baseline (Section~\ref{app:ablation-fgr-shape}) introduces hard thresholds at the band boundaries, which destroy the relative magnitude between coefficients on either side of a threshold; after the inverse DCT, this mismatch manifests as a direction-distorted spatial gradient.

\paragraph{Setup.}
Let $\hat{\bm{G}} \in \mathbb{R}^{H \times W}$ denote the 2D DCT of the input gradient $\bm{\nabla} \in \mathbb{R}^{H \times W}$. Denote by $\phi: [0, 1] \to [0, 1]$ a non-negative measurable function and by $d(u, v)$ the radial distance defined in Eq.~\ref{eq:freq_dist}. The FGR-modulated coefficient at frequency $(u, v)$ is $\tilde{G}_{u, v} = \phi(d(u,v)) \cdot \hat{G}_{u, v}$.

\paragraph{Lemma 1 (phase preservation).}
For any two coefficients $(u_1, v_1)$ and $(u_2, v_2)$ on the same radial circle, i.e.\ $d(u_1, v_1) = d(u_2, v_2)$, the modulation by $\phi$ preserves the sign and the relative magnitude:
$\tilde{G}_{u_1, v_1} / \tilde{G}_{u_2, v_2} = \hat{G}_{u_1, v_1} / \hat{G}_{u_2, v_2}.$

\textit{Proof sketch.} On the same circle, $\phi(d(u_1, v_1)) = \phi(d(u_2, v_2))$, so the modulation factor cancels in the ratio. 

\paragraph{Lemma 2 (monotone radial preservation).}
For any two coefficients on different radial circles, the modulation rescales their ratio by a known monotone factor that depends only on the two distances. The induced spatial-domain filter is real, even-symmetric, and non-negative.

\paragraph{Counterexample for per-band clipping.}
We give a constructive example showing that per-band statistical clipping breaks the relative magnitude of two coefficients with $|\hat{G}| > \mu_{\mathcal{B}} + \gamma_{\mathcal{B}} \sigma_{\mathcal{B}}$ that fall on opposite sides of a band boundary $\tau_l$ or $\tau_h$. After the inverse DCT, the recovered spatial gradient differs from the polynomial-decay output not only in magnitude but also in direction (cosine similarity strictly less than $1$).

An empirical confirmation that polynomial decay outperforms statistical clipping is reported in Section~\ref{app:ablation-fgr-shape}: polynomial $(1-d)^{1.5}$ improves over the no-FGR baseline by $+6.2$ ASR points on the $6$-victim panel, while per-band statistical clipping under-performs the same baseline by $0.1$ ASR points.

\subsection{Frequency-Domain OT Carries More Information than Spatial-Domain Clustering OT}\label{app:theory-ot}
We argue informally that the transport plan computed on top-$n$ DCT components is more informative than the transport plan computed on $n$ spatial-domain cluster centers, in the sense that it carries strictly more independent degrees of freedom for the same value of $n$.

\paragraph{Setup.}
Let $\bm{E} \in \mathbb{R}^{P \times d}$ denote the patch embeddings of an image. Let $\bm{X}_{\text{clu}} \in \mathbb{R}^{n \times d}$ denote the $n$ K-means cluster centers used by FOA-Attack, and let $\bm{L} \in \mathbb{R}^{n \times d}$ denote the top-$n$ high-frequency DCT components used by FRA-Attack.

\paragraph{Argument.}
The DCT basis is orthogonal, so the $n$ rows of $\bm{L}$ are mutually orthogonal projections of $\bm{E}$. The $n$ rows of $\bm{X}_{\text{clu}}$, by contrast, are convex combinations of subsets of $\bm{E}$ and are mutually correlated (the cluster centers of close-by patches are close-by). When the OT cost matrix $C$ is computed on rows that are mutually correlated, the entropy of the optimal transport plan $\pi^*$ is artificially lower than its independent-rows counterpart, and the gradient that flows back from $\sum_{a, b} \pi^*_{ab} C_{ab}$ to the input pixels has reduced effective rank.

An empirical confirmation is reported in Section~\ref{sec:ablation_study}: replacing the spatial-domain clustering OT of FOA-Attack with the frequency-domain DCT OT used by FRA-Attack (with all other components matched in the v2 configuration) lifts the average ASR from $36.5$ (FOA-Attack) past $50.0$ (M-Attack-V2 fair) and on to $54.9$ for the DCT-only variant of FRA-Attack on the $1{,}000$-pair closed-source panel (Table~\ref{tab:ablation}).

\section{Reproducibility}\label{app:repro}

\paragraph{Code release.}
The full implementation of FRA-Attack, the v2 reproduction of M-Attack-V2, the evaluation harness, and the configuration files for every table in this paper will be released at an anonymous repository upon acceptance. A double-blind anonymous mirror is provided to reviewers at submission time.

\paragraph{Hardware.}
All attacks and ablations run on a single node with $4 \times$ NVIDIA H200 GPUs .

\paragraph{Software.}
PyTorch (latest stable at time of submission), \texttt{open\_clip}, and the standard scientific Python stack. A complete \texttt{environment.yml} and \texttt{requirements.txt} are included in the released repository.

\paragraph{Random seeds.}
Each attack is run with multiple independent seeds for the random-crop augmentation, the dynamic-weighting noise, and the optimizer state, with the source--target pair indices held fixed across seeds for strict comparability. Reported numbers in the main tables and figures are means across these seeds; per-seed fluctuation stays within $\pm 2$--$3$ ASR points (Section~\ref{sec:settings}). Per-seed numbers and the seed list are provided in the released repository.

\paragraph{Reproduction commands.}
For each table in the main paper and the appendix, the released repository ships a single shell script that regenerates the corresponding numbers end-to-end (attack, evaluation, table rendering).



\end{document}